\documentclass[runningheads]{llncs}

\usepackage{eccv}

\usepackage{eccvabbrv}

\usepackage{graphicx}
\usepackage{booktabs}
\usepackage{colortbl}
\usepackage{xcolor}
\usepackage{multirow}
\usepackage{amsmath}
\usepackage{textcomp}

\usepackage[accsupp]{axessibility}

\usepackage{hyperref}

\usepackage{orcidlink}

\begin{document}

\title{MELON: A Large-Scale Dataset for Multi-Event Text-to-Long-Video Retrieval} 

\titlerunning{MELON}

\author{Chan Hur\inst{1}\thanks{Equal contribution.} \and
SeungWoo Song\inst{2}$^{\star}$ \and
Jeong-hun Hong\inst{3}$^{\star}$ \and
Won Jun Oh\inst{2} \and
Hyeyoung Park\inst{3} \and
KyungTae Lim\inst{2}}

\authorrunning{C. Hur et al.}

\institute{ETRI \and KAIST \and Kyungpook National University \\
\email{chanhur@etri.re.kr}, 
\email{\{sswoo, wjoh, ktlim\}@kaist.ac.kr}, 
\email{\{invhun, hypar\}@knu.ac.kr}}

\maketitle

\begin{abstract}
Existing text-video retrieval datasets primarily consist of short-form clips containing a single dominant event. While suitable for measuring basic vision–language alignment, they are limited in capturing real-world retrieval scenarios, where long-form videos naturally contain multiple semantically distinct events and a single text query may correspond to several non-contiguous temporal segments. To bridge this gap, we introduce MELON, the first large-scale dataset designed to extend text-video retrieval to long-form videos featuring complex, multi-event structures. MELON explicitly annotates multiple event intervals per video along with their corresponding textual descriptions, enabling both training and evaluation of multi-event understanding in long, untrimmed videos. In addition, we propose a multi-event aware loss that encourages models to differentiate between full-event and partial-event matches, yielding substantial improvements in retrieval accuracy. Together, the MELON dataset and our proposed loss establish a robust foundation for expanding text-to-video retrieval to complex long-form scenarios and provide a more realistic evaluation setting for future research in the field.
\end{abstract}    
\section{Introduction}
\label{sec:intro}

\begin{figure}[t] %
    \centering
    \includegraphics[width=0.8\linewidth]{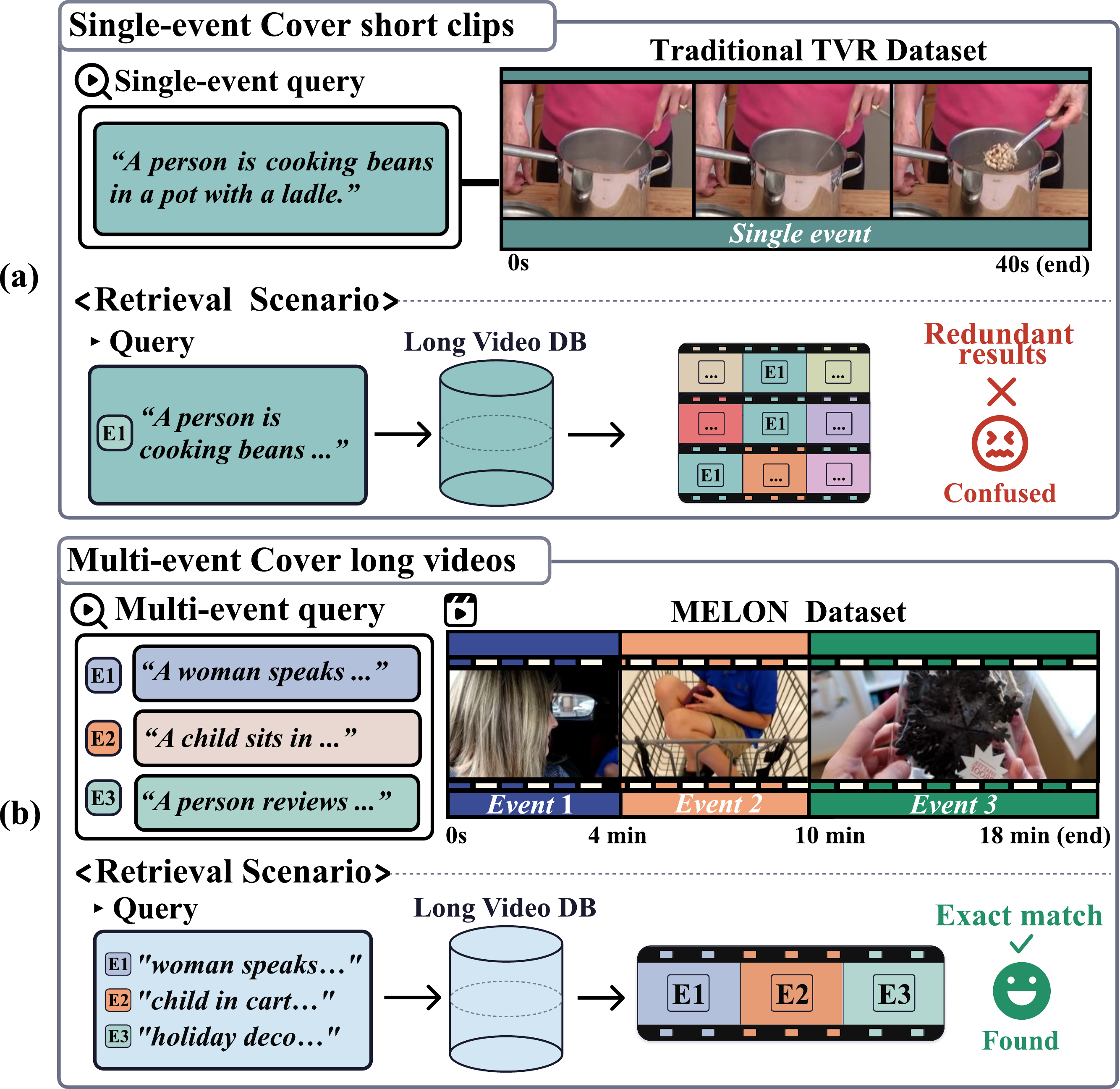} %
    \caption{Comparison of traditional text-video retrieval dataset for single-event query retrieval (a) and MELON  dataset for multi-event query retrieval (b).}
    \label{fig:teaser}
\end{figure}

Text-to-Video Retrieval (TVR) has steadily emerged as a core research problem focused on learning semantic correspondences between videos and language, with the goal of retrieving the most relevant video given a textual query. To support this task, many representative datasets, such as MSR-VTT \cite{msrvtt}, MSVD \cite{msvd}, DiDeMo \cite{didemo}, ActivityNet-Cap \cite{activitynet_captions}, and VATEX \cite{vatex}, have been introduced.

Existing datasets for TVR tasks are predominantly constructed from short-form videos that contain a single dominant event, as summarized in \cref{tab:dataset_comparison_expanded}. 
These datasets were originally designed for captioning or question-answering settings that assume a single event per video. As a result, prior work has reported that when a video contains multiple salient events, \textit{answer/annotation ambiguity} frequently arises \cite{activitynet_captions, msrvtt}.
Even benchmarks that explicitly incorporate multiple events, such as ActivityNet-Cap and DiDeMo, remain limited by their relatively short durations compared to real-world content. Consequently, while suitable for evaluating basic vision–language alignment, these existing paradigms fail to capture the full complexity of recent real-world video retrieval scenarios.

In the era of diverse video platforms, video content is increasingly expanding to long-form videos that encompass a wide range of events over a long time span. A critical but often overlooked challenge in retrieving these videos is event redundancy. As conceptually illustrated in \cref{fig:teaser}(a), traditional TVR paradigms rely on short clips paired with single-event queries, which leads to high retrieval ambiguity yielding redundant matches and leaving the model unable to discriminate between candidates. To resolve this, we argue that a comprehensive context covering multiple events is essential to filter out redundant results and pinpoint an `exact match', as shown in \cref{fig:teaser}(b). Although previous work such as MeVTR \cite{mevtr} explored multi-event retrieval scenario, it remains constrained by its foundation in short-video datasets \cite{activitynet_captions, Charades} and its focus on a one-to-many task—retrieving independent event text rather than capturing the composite narrative structure required for long-form video retrieval.

Building on these insights, we introduce MELON (\textbf{M}ulti-\textbf{E}vent Text-to-\textbf{Lon}g Video Retrieval), the first large-scale text-video retrieval dataset that extends the task to long-form videos with complex multi-event structures. MELON contains over 12K long-form videos across diverse categories and directly addresses the limitations of existing single-event–centric datasets. A central characteristic of MELON is that each long video is annotated with a set of independent event-level descriptions, where a single description may correspond to multiple distinct temporal intervals within the same video. %
The resulting dataset spans a wide variety of domains, including daily life, cooking, travel, and educational content, facilitating comprehensive evaluation of realistic video retrieval.

\begin{table}[t]
\caption{Comparison of representative video datasets used for text-to-video retrieval. }
\label{tab:dataset_comparison_expanded}
\centering
\resizebox{0.7\linewidth}{!}{%
\begin{tabular}{@{}c|c|c|c|c|c@{}}
\toprule[1.5pt]
\textbf{Dataset} & \textbf{Year} & \textbf{Avg. Duration} & \textbf{\#Video} & \textbf{\#Query/Event} & \textbf{Multi-Event} \\ 
\midrule
MSVD \cite{msvd} & 2011 & 10 $sec$ & 1.9K & 70K/- & No \\
MSR-VTT \cite{msrvtt} & 2016 & 10 $sec$ & 10K & 200K/- & No \\
LSMDC \cite{lsmdc} & 2016 & 4-5 $sec$ & 118K & 118K/- & No \\
Charades \cite{Charades} & 2016 & 30 $sec$ & 9.8K & 9.8K/- & No \\
ActivityNet-Cap \cite{activitynet_captions} & 2017 & 2 $min$ & 20K & 20K/100K & Yes \\
DiDeMo \cite{didemo} & 2017 & 30 $sec$ & 10K & 10K/40K & Yes \\
VATEX \cite{vatex} & 2019 & 15 $sec$ & 41K & 825K/- & No \\
TVshow Retrieval \cite{tvr} & 2020 & 76.2 $sec$ & 21.8K & 21.8K/109K & Yes \\
RTime \cite{rtime} & 2024 & 20.4 $sec$ & 21K & 210K/- & No \\
MultiVENT 2.0 \cite{multivent} & 2025 & 145 $sec$ & 218K & 3.9K/- & No \\
\midrule
\textbf{MELON (Ours)} & \textbf{2025} & \textbf{12.4 \textit{min}} & \textbf{12K} & \textbf{12K/42K} & \textbf{Yes} \\
\bottomrule[1.5pt]
\end{tabular}%
}

\end{table}

Furthermore, addressing the complexities of long-form video retrieval requires specialized learning strategies that account for intricate multi-event correspondences. Specifically, models must distinguish between comprehensive matches and deceptive partial overlaps. To this end, we propose the Multi-Event Aware (MEA) loss, a plug-and-play objective designed to boost a model’s discriminative power within the standard retrieval framework. By explicitly supervising the embedding space to recognize that query subsets should yield lower similarity scores than the full multi-event query, MEA encourages capturing the entire narrative context without requiring explicit temporal video segmentation. This approach consistently yields significant performance gains across various architectures and benchmarks. Our contributions are summarized as follows:
\begin{itemize}
    \item \textbf{Large-scale Long-form Multi-Event Dataset:} We construct and release MELON, the first large-scale long-form video dataset for the TVR task featuring complex, realistic multi-event scenarios, addressing the limitations of prior single-event and short-clip–oriented datasets.
    \item \textbf{Multi-Event Aware Loss:} We propose a novel Multi-Event Aware (MEA) loss specifically designed to enhance event-level discrimination and alignment in complex long-form video retrieval scenarios.
    \item \textbf{Comprehensive Empirical Validation:} Through extensive experiments across representative retrieval models, we demonstrate the difficulty, realism, and benchmarking value of the MELON dataset, and show that the proposed MEA loss integrates smoothly and consistently improves performance.
\end{itemize}

\section{Related works}
\label{sec:related_works}

\textbf{Datasets for Text–Video Retrieval.} Progress in TVR has been largely enabled by a series of influential datasets such as MSVD \cite{msvd}, MSR-VTT \cite{msrvtt}, and VATEX \cite{vatex}, whose core statistics are summarized in \cref{tab:dataset_comparison_expanded}. 
While foundational, these datasets predominantly feature short clips (averaging 5–15 seconds) where a single-event query corresponds to the entire video. Although ActivityNet-Cap \cite{activitynet_captions} and DiDeMo \cite{didemo} incorporate multiple events, their temporal scale remains limited, failing to reflect the complexity of real-world long-form content.
A number of domain-specific datasets have also been adopted for TVR. LSMDC \cite{lsmdc} includes short segments extracted from movies with an average duration of 4–5 seconds, while the recently released MultiVENT 2.0 \cite{multivent} provides news videos averaging 145 seconds, they focus on single event.
Despite their utility, most existing TVR datasets fall short in capturing the semantic complexity, event diversity, and extended temporal dynamics inherent in the multi-event, long-form scenarios.

\noindent\textbf{Text-Video Retrieval Methods.} Given the availability of text–video paired datasets, recent TVR approaches \cite{clip4clip, xclip, ucofia, uatvr, pau, ts2net, cap4video, tefal, narvid, avigate, tempme} have advanced rapidly by leveraging large-scale vision–language pre-trained models such as CLIP \cite{clip}. CLIP4Clip \cite{clip4clip} extends CLIP to the video domain via frame-level feature aggregation. Subsequent methods—including X-CLIP \cite{xclip}, TS2Net \cite{ts2net}, and UCOFIA \cite{ucofia}—achieve improved performance by modeling fine-grained cross-modal interactions. More recent efforts, such as UATVR \cite{uatvr} and PAU \cite{pau}, incorporate uncertainty modeling to better capture inherent ambiguities in video–text correspondence. Additionally, TempMe \cite{tempme} enhances efficiency by merging temporal tokens to reduce spatio-temporal redundancy. These methods largely rely on contrastive learning (\eg, InfoNCE \cite{infonce}) to align video and text representations in a joint embedding space. While this paradigm works well for conventional short-video datasets containing a single dominant event, it becomes suboptimal when applied to long, untrimmed videos composed of multiple semantically distinct events. 
Recently, MeVTR \cite{mevtr} addressed a one-to-many problem where a video retrieves independent event-level texts. However, focusing on partial alignments limits its ability to capture the holistic relationships and broader narrative structures essential for multi-event, long-video scenarios.

\section{The MELON Dataset}

In this section, we detail the motivation and the automated pipeline for constructing MELON (\textbf{M}ulti-\textbf{E}vent Text-to-\textbf{Lon}g Video Retrieval), a large-scale dataset designed to scale text-to-video retrieval to realistic long-form settings featuring complex multi-event structures. Traditionally, the construction of long-form video datasets with multiple annotations has been severely constrained by prohibitive annotation costs and procedural complexity. Because long-form videos inherently contain sequences of continuous events, fine-grained temporal validation is highly challenging, which limits the ability to ensure model generalization across diverse domains \cite{activitynet_captions}. To overcome these barriers, we design an automated data generation pipeline inspired by efficient strategies that repurpose existing data sources to construct high-quality benchmarks at a low cost \cite{Brightsun, WILD, vist, vid2seq, chapterllama}.

\begin{figure*}[t]
    \centering
    \includegraphics[width=\textwidth]{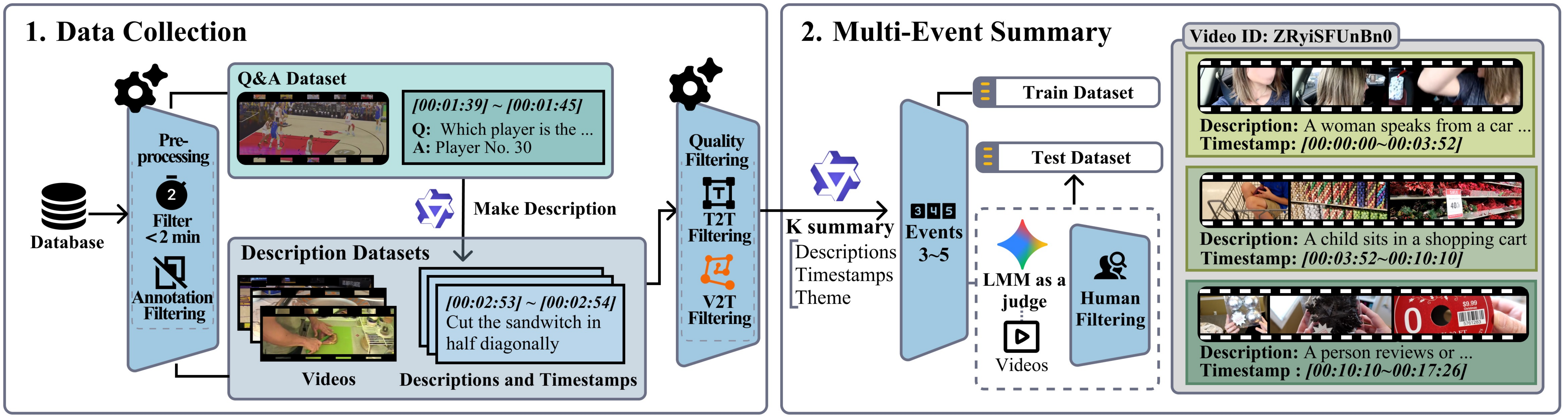}

    \caption{Overview of the MELON data pipeline. 
    \textbf{(Left) Data Collection:} Long videos are pre-processed and annotations unified. Quality filtering subsequently then removes event-redundant videos (Text-to-Text) and misaligned timestamps (Video-to-Text). 
    \textbf{(Right) Multi-Event Summary:} An LLM generates 3-5 salient event summaries. Test data undergoes further validation by an LMM (judge) and human filtering.}
    
    \label{fig:dataset_pipeline}
\end{figure*}

As illustrated in \cref{fig:dataset_pipeline}, this pipeline involves a multi-stage process that leverages the reasoning capabilities of Qwen3-235B-A22B \cite{qwen3technicalreport} for data refinement, filtering, and event-level synthesis. MELON, built through this pipeline, is specifically designed to evaluate text-video retrieval performance in realistic, extended temporal settings. It directly addresses the core challenge of multi-event retrieval by curating videos that are at least 2 minutes long, with each containing 3 to 5 non-overlapping events. The resulting dataset comprises 12,000 samples (8,400 for training and 3,600 for testing), faithfully reflecting real-world retrieval scenarios where complex events unfold over long durations.

\noindent\textbf{Data Acquisition and Pre-processing.} To construct a high-quality dataset targeting complex event understanding, we curated long-form video data by identifying six recent and representative benchmarks~\cite{cgbench, youcook2, longvale, scenewalk, vidchapter7m, vrbench}. These sources were specifically selected to reflect the latest advancements in video-language tasks, as they feature extended durations and dense multi-event structures that surpass the limitations of traditional, short-clip datasets. These datasets were strategically chosen because they: (1) consist of long-form content that aligns with our focus on capturing extended temporal dependencies, and (2) provide fine-grained, valid timestamp annotations essential for multi-event localization. We applied a rigorous pre-processing pipeline, as illustrated in the ``Data Collection'' stage of \cref{fig:dataset_pipeline}. This process included discarding videos shorter than 2 minutes to ensure temporal depth, retaining only samples with high-fidelity and comprehensive timestamps, and filtering out descriptions with insufficient content (\eg, event-sparse or misaligned annotations).

\noindent\textbf{Data Merging and Annotation Refinement.}
Next, we perform a multi-stage merging and refinement process (\cref{fig:dataset_pipeline}, left) to construct a unified resource with consistent multi-event temporal annotations. After merging the curated sources, we obtain event-level descriptions by adapting annotation formats from each dataset. If a dataset provides dense, high-quality temporal captions, we directly adopt these annotations. For datasets offering timestamped question–answer (QA) pairs, we employ an LLM to convert each QA pair into a detailed caption that accurately reflects the corresponding time interval. The specific prompting strategy and examples are provided in Supplementary E. This dual-path strategy significantly increases the volume and richness of event descriptions while preserving alignment with the underlying temporal structure of long videos.  Finally, we consolidate the refined annotations into a structured set of $M$ event samples:
$\{(Description_i, Timestamp_i)\}_{i=1}^{M}$.

\noindent\textbf{Quality Assurance and Filtering.} \label{sec:quality_filtering} After annotation refinement, each video contains multiple \text{Description}, \text{Timestamp} pairs. However, the number and quality of these annotations vary considerably across the source datasets, requiring further selection to ensure suitability for multi-event retrieval. In particular, we observe two common issues:
(1) some videos do not contain multiple meaningful events (\eg, continuous scenes such as a static fireplace), and (2) certain descriptions poorly correspond to the visual content. To address these issues, we employ two robust filtering mechanisms, as illustrated on the right side of the ``Data Collection'' block in \cref{fig:dataset_pipeline}.

\begin{itemize}
\item T2T (Text-to-Text matching): We discard videos if the similarity between consecutive descriptions is above a certain score, indicating event redundancy. (T2T similarity $ > 0.70 $)
\item V2T (Video-to-Text matching): We remove individual timestamps if the similarity between the video and text is below a dataset-specific threshold. (V2T similarity $\approx 0.12$–$0.18$)
\end{itemize}

The T2T filtering step identifies videos lacking genuine multi-event structure by detecting high redundancy between adjacent descriptions; 14.2\% of videos are removed at this stage. Next, V2T filtering eliminates low-quality timestamp annotations, removing approximately 16.6\% of the event segments. Detailed descriptions of all filtering parameters, including T2T thresholds and dataset-wise V2T statistics, are provided in Supplementary C.

\noindent\textbf{Multi-Event Summary Generation.} Even after filtering, the remaining annotations constitute a dense set of fine-grained segments ($M'$ pairs of $Description$, $Timestamp$), which are too detailed to serve directly as ground truth for retrieval. Using the full set would lead to numerous trivial matches, exacerbate retrieval noise, and hinder meaningful evaluation. Thus, to design a realistic \emph{multi-event} retrieval dataset, we must condense these dense segments into a small set of $K$ salient events that capture the core narrative of each video. We therefore perform a multi-event summary process that produces $K$ representative event summaries per video, where $K$ ranges from 3 to 5. This step steers the retrieval task toward major, semantically meaningful events rather than fragment-level details. As shown in the ``Multi-Event Summary'' stage of \cref{fig:dataset_pipeline}, we leverage the Qwen3 model to extract and synthesize concise, high-quality summaries representing the video’s key events (see Supplementary E for the prompting strategy). Since the target application is retrieval, we avoid using excessively large $K$, which would distort natural search-query behavior and introduce unnecessary annotation noise. Each final event summary consists of three essential components: $Theme$, $Description$, and $Timestamp$.

\noindent\textbf{Human Filtering for Test Data.} To rigorously ensure the reliability of the test set, we introduce an additional validation stage that combines multimodal LMM screening with human review. Specifically, we utilize a multimodal LMM (Gemini~\cite{gemini-flash}) to automatically evaluate the consistency between each summary's $Description$, $Timestamp$, and the corresponding video content, thereby identifying potential alignment errors. Through this robust screening followed by human verification, we guarantee that the final test set maintains a high level of annotation fidelity. Detailed statistics of the filtering process and the guidelines for the human review step are provided in Supplementary C.

\begin{figure}[t]
    \centering
    \includegraphics[width=\textwidth]{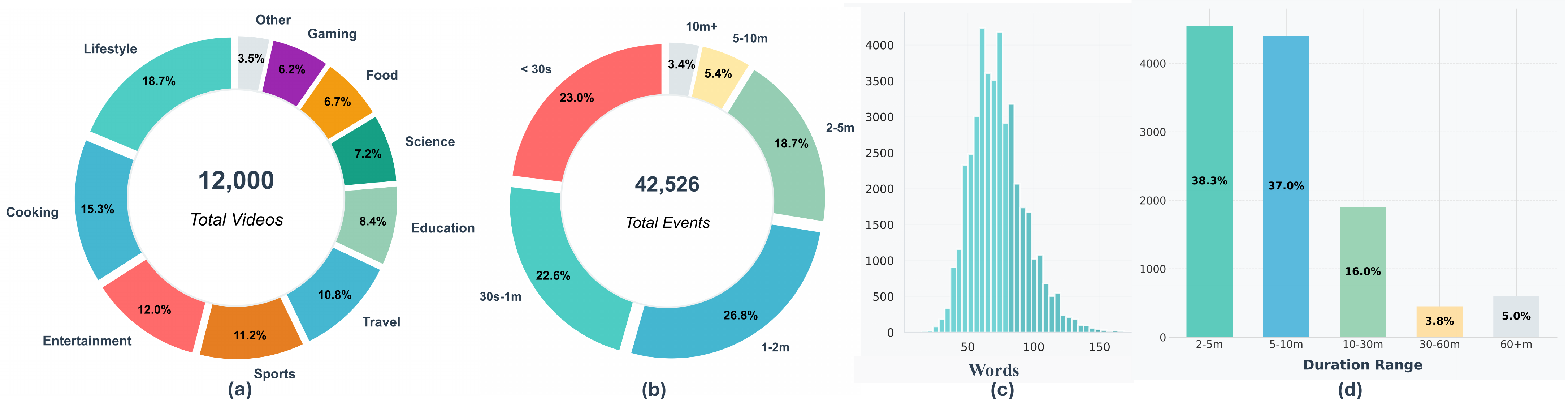}

    \caption{Quantitative statistics of the MELON dataset. 
    (a) Thematic distribution of 12,000 videos across themes. 
    (b) Temporal duration distribution of 42,526 annotated events. 
    (c) Distribution of description lengths by word count.
    (d) Distribution of source video durations. }
    \label{fig:dataset_stats} 
\end{figure}

\noindent\textbf{Dataset Statistics.} MELON comprises 12,000 long-form videos and 42,526 annotated events. As shown in \cref{fig:dataset_stats}(a), the data is balanced across ten semantic themes (\eg, Food, Lifestyle, Sports) for domain diversity. Annotated events range from 30 seconds to multi-minute intervals (\cref{fig:dataset_stats}(b)), with descriptions averaging 70–90 words (\cref{fig:dataset_stats}(c)). Notably, over 75\% of videos span 2–10 minutes, and 5.0\% exceed 60 minutes (\cref{fig:dataset_stats}(d)), emphasizing the need for capturing long-range temporal dependencies. See Supplementary D for per-category density and linguistic statistics.
\section{Method}
\label{sec:method}

\begin{figure}[t]
    \centering
    \includegraphics[width=0.75\columnwidth]{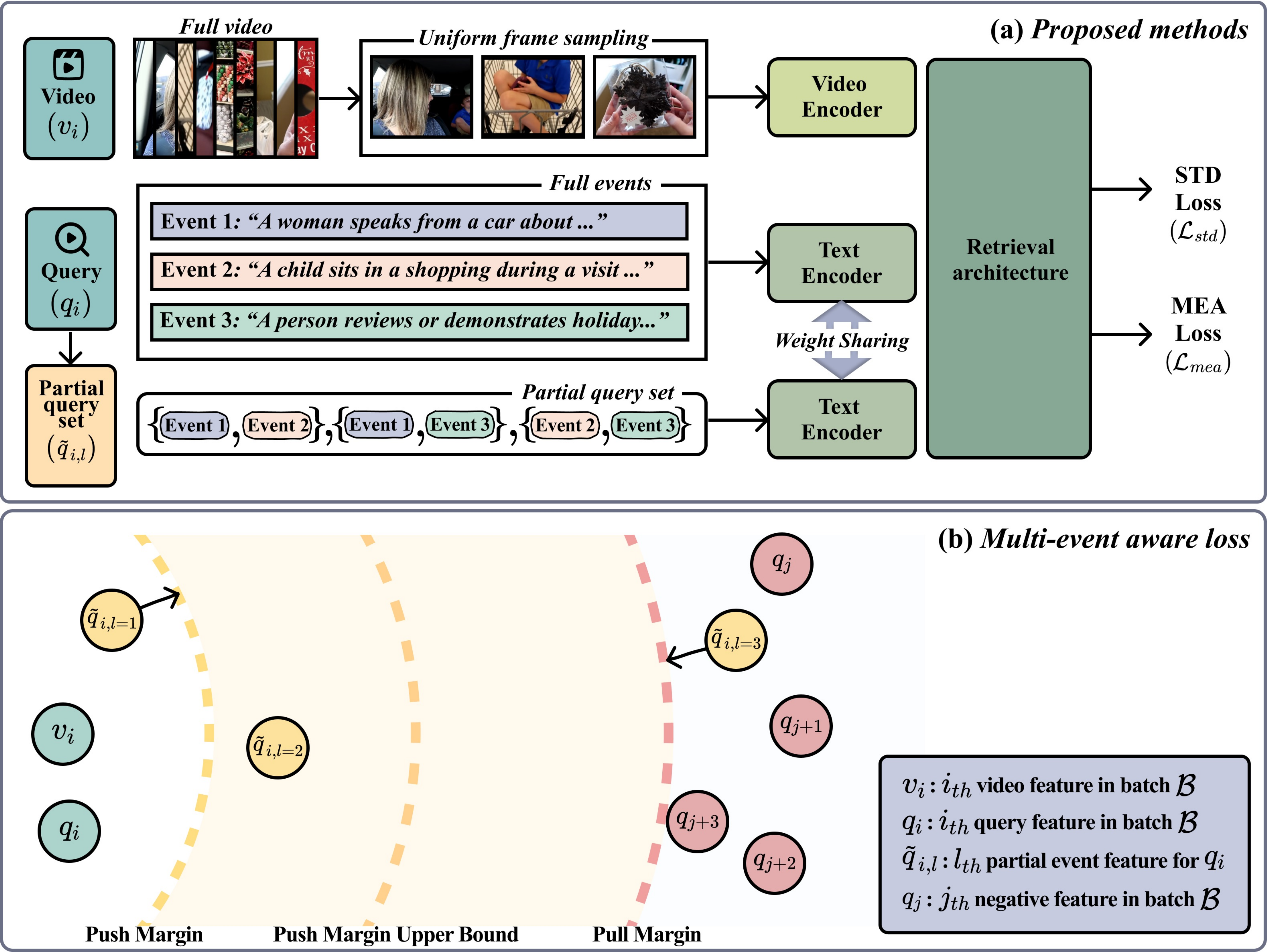}
    \caption{
        (a) Overview of the proposed method.
        (b) Multi-event aware loss: encourages full query–video matching and improves partial alignment via push–pull optimization.
    }
    \label{fig:method}
\end{figure}

\subsection{Problem Definition}

MELON is designed to overcome the limitations of existing retrieval datasets that rely on single-event or short-clip queries. In our setting, a retrieval query is defined as a set of $K$ event-level descriptions, queries: $q=\{e_{1},e_{2},...,e_{K}\}$, where each $e_k$ is a natural-language sentence describing a distinct event.
The corresponding video $v$ consists of $T$ frames, $v=\{f_{1},f_{2},...,f_{T}\}$ and may contain multiple semantically distinct events across extended temporal spans.

A long-video retrieval model must determine how well the video $v$ matches the multi-event query $q$. We define the retrieval score as:
\begin{equation}
\text{score} = \text{RetrievalModel}(v, q).
\label{eq:retrieval}
\end{equation}
The objective of multi-event TVR is to retrieve a video that encompasses all relevant events within the video for a given multi-event query.

\subsection{Base-Retrieval Model Architecture}

As illustrated in \cref{fig:method}(a), TVR follows a dual-encoder paradigm consisting of three essential components: a video encoder, a text encoder, and a score function.

\begin{itemize}
\item \textbf{Video encoder:} The full video is uniformly sampled into $T$ frames, which are encoded to produce a compact video representation capturing both visual and temporal cues.
\item \textbf{Text encoder:} The input tokens of the query are transformed into a text representation.
\item \textbf{Scoring function:} A similarity function computes the correspondence between the video and text.
\end{itemize}

This dual-encoder architecture underlies many existing retrieval models \cite{clip4clip, xclip, ts2net, ucofia, uatvr, pau, tempme}. In our evaluation, we apply these models directly to MELON. For multi-event queries, we concatenate the $K$ event descriptions into a single textual input before encoding, enabling baseline models to process multi-event information without architectural modification.

\subsection{Loss Design for Multi-Event Scenarios}
\subsubsection{Standard Contrastive Learning.}

Most existing TVR models are trained using a standard contrastive learning objective. This loss encourages high similarity for the correct video–query pair $(v_i, q_i)$ in a batch $\mathcal{B} = { (v_i, q_i) }_{i=1...B}$, while pushing down the similarity between mismatched pairs $(v_i, q_j)$ where $i \neq j$. The widely used InfoNCE loss \cite{infonce}, denoted as $\mathcal{L}_{\text{std}}$, is defined as:
\begin{equation}
\begin{split}
\mathcal{L}_{q2v} &= 
-\frac{1}{B} \sum_{i=1}^{B}
\log \left(
\frac{\exp(s(v_i, q_i) / \tau)}
{\sum_{j=1}^{B} \exp(s(v_j, q_i) / \tau)}
\right), \\
\mathcal{L}_{v2q} &=
-\frac{1}{B} \sum_{i=1}^{B}
\log \left(
\frac{\exp(s(v_i, q_i) / \tau)}
{\sum_{j=1}^{B} \exp(s(v_i, q_j) / \tau)}
\right), \\
\mathcal{L}_{std} &= \mathcal{L}_{q2v} + \mathcal{L}_{v2q},
\end{split}
\label{eq:std}
\end{equation}
where $\tau$ is a temperature parameter and $s(\cdot, \cdot)$ computes the similarity between video and text representations produced by their respective encoders.

Although $\mathcal{L}_{\text{std}}$ effectively aligns video and text in conventional settings, it becomes insufficient in multi-event retrieval. Consider a video annotated with three distinct events: an ideal match must contain all three. However, a partially relevant video containing only two of these events may still exhibit high similarity to the query. This issue becomes pervasive in long-form videos, where many samples naturally share subsets of events. As a result, models trained with standard contrastive loss tend to overestimate the relevance of such partially overlapping videos. We refer to this as the \textbf{partial event matching problem}, which leads to incorrect retrieval rankings in realistic multi-event scenarios.

\noindent\textbf{Multi-Event Aware Loss.}
We begin with a fundamental question: \textit{``Can we distinguish partially overlapped situations in multi-event video retrieval?''} To mitigate the partial-event matching problem and to encourage complete event coverage during training, we introduce a novel MEA loss.

\noindent\textbf{Partial Query Set.}
A partial query, constructed from a subset of the full event query, should reasonably exhibit a similarity to the corresponding video that is (1) higher than that of an unrelated query, yet (2) lower than that of the full multi-event query, since it does not cover all events. %
Inspired by this idea, we propose the use of partial queries for MEA loss construction. A partial query $\tilde{q}_{i,l}$ is formed by concatenating event sentences that constitute a proper subset of the original query $q_i$'s constituent event sentences. These queries are generated by systematically forming subsets of event sentences from $q_i$ such that a predefined number of events, $m$, are omitted. From the pool of candidates generated this way, $L$ distinct partial queries, denoted as $\{\tilde{q}_{i,l}\}_{l=1}^{L}$, are randomly selected and utilized during training. Each such $\tilde{q}_{i,l}$ still pertains to a portion of the events depicted in video $v_i$, and is thus inherently expected to exhibit a higher semantic similarity to $v_i$ compared to fully unrelated text queries $q_j$ from other videos ($j \neq i$). Consequently, these partial queries should be located near the decision boundary between positive and negative pairs. %

\noindent\textbf{Margin Loss with Adaptive Threshold (push loss).} Leveraging this set of partial queries, we design a margin-based \cite{marginloss} pushing loss $\mathcal{L}_{push}$ that regulates the similarity between the original video–query pair $s(v_i, q_i)$ and the video-partial query $s(v_i, \tilde{q}_{i,l})$.
The objective is to enforce that the similarity score $s(v_i, \tilde{q}_{i,l})$, which reflects a partial-match relationship, remains lower than that of the original positive pair by at least the adaptive threshold $\theta^t$. This explicit differentiation is crucial for the model to distinguish between $q_i$ and $\tilde{q}_{i,l}$, thereby learning the compositional differences among events.
Using the $\theta^t$, the pushing loss is expressed as:
\begin{equation}
\mathcal{L}_{push} = \frac{1}{BL}\sum_{i=1}^{B} \sum_{l=1}^{L} \max \left( 0, s(v_i, \tilde{q}_{i,l}) - s(v_i, q_i) + \theta^t \right).
\end{equation}
Static margins often lack adaptation to the model's evolving discriminative capacity due to their fixed nature. To stabilize training, we adapt a curriculum learning approach with a dynamic threshold \cite{cmrloss, adaptive_th1, adaptive_th2, adaptive_th3, adaptive_th4}. We gradually adjust the threshold $\delta^t$, which is computed based on the similarity gap from the previous training step $t - 1$:
\begin{equation}
\delta^t = \frac{1}{BL} \sum_{i=1}^{B} \sum_{l=1}^{L} \left( s^{t-1}(v_i, q_i) - s^{t-1}(v_i, \tilde{q}_{i,l}) \right),
\end{equation}
where $s^{t-1}(\cdot, \cdot)$ denotes the similarity score computed in the previous training step.  
To prevent partial queries from being excessively penalized, we further introduce an upper bound $\xi^t$ which is derived from half the average similarity gap between positive pairs and their hardest in-batch negative pairs:
\begin{equation}
\xi^t = \frac{1}{2B} \sum_{i=1}^{B} \max \left( 0, s^{t-1}(v_i, q_i) - \max_{j \neq i} s^{t-1}(v_i, q_j) \right).
\end{equation}
Finally, the adaptive threshold at step $t$ is defined as:
\begin{equation}
\theta^t = \min (\delta^t, \xi^t).
\end{equation}
\noindent\textbf{Pull Loss.} In addition, we introduce a pull loss $\mathcal{L}_{pull}$ to prevent the similarity of partial-match pairs from falling below the similarity of the hardest negative pair $(v_i, q_j)$ in the batch ($j \neq i$):
\begin{equation}
\mathcal{L}_{pull} = \frac{1}{BL}\sum_{i=1}^{B} \sum_{l=1}^{L} \max \left( 0, \max_{j \neq i} s(v_i, q_j) - s(v_i, \tilde{q}_{i,l}) \right)
\label{eq:pull2}.
\end{equation}
This pull margin is visually represented in \cref{fig:method}(b) by the red dashed line.
\noindent\textbf{Total Loss.}
The proposed MEA loss $\mathcal{L}_{mea}$ is defined as:
\begin{equation}
\mathcal{L}_{mea}
= 
\mathcal{L}_{push}
+ \mathcal{L}_{pull}.
\end{equation}
To preserve fundamental text–video alignment capability, we also incorporate the standard contrastive loss $\mathcal{L}_{\text{std}}$. The final training objective is defined as:
\begin{equation}
\mathcal{L}_{total}
=
\mathcal{L}_{std}
+
\lambda \cdot \mathcal{L}_{mea},
\end{equation}
where $\lambda$ is a weighting hyperparameter. While $\mathcal{L}_{std}$ enforces general alignment between the paired video $v$ and query $q$, the MEA loss $\mathcal{L}_{mea}$ improves the model’s discriminative capability in multi-event scenarios by explicitly distinguishing partial event queries.

\definecolor{lightgray}{gray}{0.95}
\definecolor{darkgray}{gray}{0.7}
\newcommand{\venue}[1]{{\scriptsize \textcolor{gray}{#1}}}

\begin{table*}[t]
\caption{Retrieval performance on the MELON dataset comparing standard methods (with/without MEA loss) and a multi-event aware method. Note that for the multi-event aware method, MEA loss is excluded as it is incompatible with the divergent one-to-many objective. Values in parentheses represent the performance difference over the respective baseline.}
\label{tab:performance_difference}
\centering
\resizebox{\textwidth}{!}{
\begin{tabular}{ll cccccc  cccccc}
\toprule[1.5pt]
\multicolumn{1}{l}{\multirow{2}{*}{\textbf{Method}}} &
  \multirow{2}{*}{} &
  \multicolumn{6}{c}{\textbf{Text-to-Video Retrieval (T2V)}} &
  \multicolumn{6}{c}{\textbf{Video-to-Text Retrieval (V2T)}} \\
\cmidrule(l){3-8} \cmidrule(l){9-14}
\multicolumn{1}{l}{} &
  \multicolumn{1}{l}{} &
  R@1 $\uparrow$ &
  R@5 $\uparrow$ &
  R@10 $\uparrow$ &
  Rsum $\uparrow$ & 
  MdR $\downarrow$ &
  MnR $\downarrow$ &
  R@1 $\uparrow$ &
  R@5 $\uparrow$ &
  R@10 $\uparrow$ &
  Rsum $\uparrow$ &
  MdR $\downarrow$ & 
  MnR $\downarrow$ \\
\midrule
\rowcolor{darkgray} \multicolumn{14}{l}{\textit{Standard methods on MELON}} \\
\multicolumn{2}{l|}{CLIP4Clip \cite{clip4clip} \venue{Neurocom'22}} &
  52.7 &
  77.1 &
  84.7 &
  214.5 &
  1.0 &
  \multicolumn{1}{c|}{{10.7}} &
  53.1 &
  76.9 &
  85.0 &
  215.0 & 
  1.0 & 
  {10.4} \\
\rowcolor{lightgray} \multicolumn{2}{r|}{\textbf{+$\mathcal{L}_{mea}$}} &
  \textbf{55.1} (\textcolor{red}{$\Delta 2.4$}) &
  \textbf{78.1} (\textcolor{red}{$\Delta 1.0$}) &
  \textbf{86.0} (\textcolor{red}{$\Delta 1.3$}) &
  \textbf{219.2} (\textcolor{red}{$\Delta 4.7$})&
  \textbf{1.0} &
  \multicolumn{1}{c|}{\textbf{10.7}} &
  \textbf{54.9} (\textcolor{red}{$\Delta 1.8$}) &
  \textbf{79.3} (\textcolor{red}{$\Delta 2.4$}) &
  \textbf{86.9} (\textcolor{red}{$\Delta 1.9$}) &
  \textbf{221.1} (\textcolor{red}{$\Delta 6.1$}) &
  \textbf{1.0} &
  \textbf{9.9} \\
  \midrule
\multicolumn{2}{l|}{X-CLIP \cite{xclip} \venue{MM'22}} &
  50.8 &
  75.8 &
  83.7 &
  210.3 &
  1.0 &
  \multicolumn{1}{c|}{{10.6}} &
  50.6 &
  76.1 &
  84.4 &
  211.1 & 
  1.0 & 
  {9.7} \\
\rowcolor{lightgray} \multicolumn{2}{r|}{\textbf{+$\mathcal{L}_{mea}$}} &
  \textbf{56.4} (\textcolor{red}{$\Delta 5.6$}) &
  \textbf{79.9} (\textcolor{red}{$\Delta 4.1$}) &
  \textbf{87.0} (\textcolor{red}{$\Delta 3.3$}) &
  \textbf{223.3} (\textcolor{red}{$\Delta 13.0$})&
  \textbf{1.0} &
  \multicolumn{1}{c|}{\textbf{9.9}} &
  \textbf{58.1} (\textcolor{red}{$\Delta 7.5$}) &
  \textbf{81.7} (\textcolor{red}{$\Delta 5.6$}) &
  \textbf{88.6} (\textcolor{red}{$\Delta 4.2$}) &
  \textbf{228.4} (\textcolor{red}{$\Delta 17.3$}) &
  \textbf{1.0} &
  \textbf{7.3} \\
  \midrule
\multicolumn{2}{l|}{TS2-Net \cite{ts2net} \venue{ECCV'22}} &
  47.6 &
  74.3 &
  \textbf{83.6} &
  205.5 &
  2.0 &
  \multicolumn{1}{c|}{\textbf{11.2}} &
  53.3 &
  76.8 &
  85.2 &
  215.3 &
  1.0 &
  9.5 \\
\rowcolor{lightgray} \multicolumn{2}{r|}{+$\mathcal{L}_{mea}$} &
  \textbf{48.1} (\textcolor{red}{$\Delta 0.5$}) &
  \textbf{74.7} (\textcolor{red}{$\Delta 0.4$}) &
   83.4 (\textcolor{blue}{$\nabla 0.2$}) &
  \textbf{206.2} (\textcolor{red}{$\Delta 0.7$}) & %
  \textbf{2.0} &
  \multicolumn{1}{c|}{11.8} &
  \textbf{54.8} (\textcolor{red}{$\Delta 1.5$}) &
  \textbf{79.1} (\textcolor{red}{$\Delta 2.3$}) &
  \textbf{86.7} (\textcolor{red}{$\Delta 1.5$}) &
  \textbf{220.6} (\textcolor{red}{$\Delta 5.3$}) &
  \textbf{1.0} &
  \textbf{9.2} \\
  \midrule
\multicolumn{2}{l|}{UCOFIA \cite{ucofia} \venue{ICCV'23}} &
 44.5  &
 70.0  &
 79.7  &
 194.2  &
 2.0  &
  \multicolumn{1}{c|}{\textbf{13.9}} &
 42.8  &
 69.1  &
 79.0  &
 190.9  &
 2.0  &
 12.4 \\
\rowcolor{lightgray} \multicolumn{2}{r|}{+$\mathcal{L}_{mea}$} &
  \textbf{47.3} (\textcolor{red}{$\Delta 2.8$}) &
  \textbf{73.8} (\textcolor{red}{$\Delta 3.8$}) &
  \textbf{82.3} (\textcolor{red}{$\Delta 2.6$}) &
  \textbf{203.4} (\textcolor{red}{$\Delta 9.2$}) &
  \textbf{2.0} &
  \multicolumn{1}{c|}{14.6} &
  \textbf{49.2} (\textcolor{red}{$\Delta 6.4$}) &
  \textbf{75.4} (\textcolor{red}{$\Delta 6.3$}) &
  \textbf{83.8} (\textcolor{red}{$\Delta 4.8$}) &
  \textbf{208.4} (\textcolor{red}{$\Delta 17.5$}) &
  \textbf{2.0} &
  \textbf{9.3} \\
  \midrule
\multicolumn{2}{l|}{UATVR \cite{uatvr} \venue{ICCV'23}} &
  52.8 &
  77.7 &
  85.8 &
  216.3 &
  1.0 &
  \multicolumn{1}{c|}{ \textbf{9.5}} &
  54.3 &
  78.6 &
  86.9 &
  219.8 &
  1.0 &
  \textbf{8.8} \\
\rowcolor{lightgray} \multicolumn{2}{r|}{\textbf{+$\mathcal{L}_{mea}$}} &
  \textbf{53.0} (\textcolor{red}{$\Delta 0.2$}) &
  \textbf{78.1} (\textcolor{red}{$\Delta 0.4$}) &
  \textbf{86.4} (\textcolor{red}{$\Delta 0.6$}) &
  \textbf{217.5} (\textcolor{red}{$\Delta 1.2$})&
  \textbf{1.0} &
  \multicolumn{1}{c|}{11.7} &
  \textbf{55.1} (\textcolor{red}{$\Delta 0.8$}) &
  \textbf{79.4} (\textcolor{red}{$\Delta 0.8$}) &
  \textbf{87.4} (\textcolor{red}{$\Delta 0.5$}) &
  \textbf{221.9} (\textcolor{red}{$\Delta 2.1$}) &
  \textbf{1.0} &
  8.9 \\
  \midrule
\multicolumn{2}{l|}{PAU \cite{pau} \venue{NeurIPS'23}} &
  48.5 &
  75.1 &
  83.0 &
  206.6 &
  2.0 &
  \multicolumn{1}{c|}{11.2} &
  50.3 &
  75.8 &
  84.2 &
  210.3 &
  1.0 &
  10.4 \\
\rowcolor{lightgray} \multicolumn{2}{r|}{+$\mathcal{L}_{mea}$} &
  \textbf{55.4} (\textcolor{red}{$\Delta 6.9$}) &
  \textbf{79.0} (\textcolor{red}{$\Delta 3.9$}) &
  \textbf{86.6} (\textcolor{red}{$\Delta 3.6$}) &
  \textbf{221.0} (\textcolor{red}{$\Delta 14.4$}) &
  \textbf{1.0} &
  \multicolumn{1}{c|}{\textbf{9.0}} &
  \textbf{55.2} (\textcolor{red}{$\Delta 4.9$}) &
  \textbf{79.6} (\textcolor{red}{$\Delta 3.8$}) &
  \textbf{87.6} (\textcolor{red}{$\Delta 3.4$}) &
  \textbf{222.4} (\textcolor{red}{$\Delta 12.1$}) &
  \textbf{1.0} &
  \textbf{7.7} \\
  \midrule
\multicolumn{2}{l|}{TempMe \cite{tempme} \venue{ICLR'25}} &
  57.6 &
  79.7 &
  86.2 &
  223.5 &
  1.0 &
  \multicolumn{1}{c|}{\textbf{9.2}} &
  56.9 &
  79.4 &
  86.2 &
  222.5 &
  1.0 &
  8.7 \\
\rowcolor{lightgray} \multicolumn{2}{r|}{+$\mathcal{L}_{mea}$} &
  \textbf{59.2} (\textcolor{red}{$\Delta 1.6$}) &
  \textbf{80.5} (\textcolor{red}{$\Delta 0.8$}) &
  \textbf{87.1} (\textcolor{red}{$\Delta 0.9$}) &
  \textbf{226.8} (\textcolor{red}{$\Delta 3.3$}) &
  \textbf{1.0} &
  \multicolumn{1}{c|}{9.6} &
  \textbf{57.4} (\textcolor{red}{$\Delta 0.5$}) &
  \textbf{80.3} (\textcolor{red}{$\Delta 0.9$}) &
  \textbf{87.5} (\textcolor{red}{$\Delta 1.3$}) &
  \textbf{225.2} (\textcolor{red}{$\Delta 2.7$}) &
  \textbf{1.0} &
  \textbf{8.0} \\
\midrule
\rowcolor{darkgray} \multicolumn{14}{l}{\textit{Multi-event aware method on MELON}} \\
\multicolumn{2}{l|}{MeVTR \cite{mevtr} \venue{ICCV'23}} &
  41.4 &
  63.2 &
  73.3 &
  177.9 &
  2.0 &
  \multicolumn{1}{c|}{12.8} &
  44.0 &
  69.1 &
  77.2 &
  190.3 & 
  2.0 & 
  18.1 \\
\bottomrule[1.5pt]
\end{tabular}
}
\end{table*}

\section{Experimental results}
\label{sec:experiment}
\subsection{Experiment Details}

All experiments are conducted and evaluated on the MELON dataset using its train/test split. We assess retrieval performance with standard metrics: Recall at K (R@K) for $\text{K}=\{1, 5, 10\}$, Median Rank (MdR), Mean Rank (MnR), and Rsum, where Rsum represents the aggregated sum of R@K scores across these ranks. Higher values of R@K and Rsum indicate better performance, whereas lower MdR and MnR values correspond to better results.

For our evaluation, we adopt seven representative CLIP-based \cite{clip} text–video retrieval models \cite{clip4clip, xclip, ts2net, ucofia, uatvr, pau, tempme}, all of which operate without external pre-trained models or additional modality-specific supervision. To adapt these architectures to the multi-event text-to-long-video setting, we replace each model’s original encoder with Long-CLIP-B/16 \cite{longclip}, which extends CLIP's text modeling capability. During training, we uniformly sample 64 frames per video (32 frames for TS2Net \cite{ts2net}). For the proposed $\mathcal{L}_{\text{mea}}$, the weighting factor $\lambda$ is tuned individually for each model to maximize performance. Detailed configurations are provided in Supplementary B.

\begin{figure}[t]
    \centering
    \includegraphics[width=1.0\linewidth]{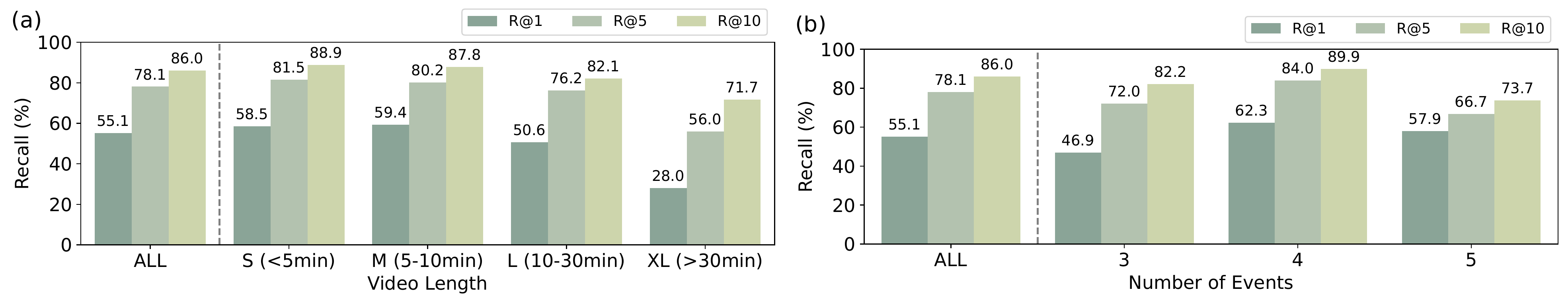}
    \caption{
        Text-to-Video Retrieval performance by video length and event count.
    }
    \label{fig:timestamp_event_comparison}
\end{figure}
\subsection{Retrieval results on MELON dataset}

\cref{tab:performance_difference} presents the performance comparison between models trained with only the standard contrastive loss and those trained with our proposed MEA loss.

\noindent\textbf{Baseline Performance on Multi-Event Retrieval.}
\cref{tab:performance_difference} highlights the intrinsic difficulty of the MELON dataset, which requires models to align complex multi-event queries with long, untrimmed videos. Most standard one-to-one models achieve around 50\% R@1, demonstrating the challenging nature of the task. Interestingly, many sophisticated architectures perform on par with—or even slightly below—the simplest baseline, CLIP4Clip \cite{clip4clip}, suggesting they struggle with the richer temporal structures of long videos. In addition to these standard baselines, we compare our setting with MeVTR \cite{mevtr}, which focuses on a one-to-many task retrieving independent single-event texts. To evaluate MeVTR in our one-to-one scenario, we perform mean pooling over its event-level embeddings. The results show that MeVTR underperforms standard one-to-one baselines, confirming that models optimized for discrete event matching are less effective at capturing the holistic narrative required for the one-to-one retrieval task.

\noindent\textbf{Performance Improvement with MEA Loss.} Introducing MEA loss consistently boosts Recall and Rsum metrics across all architectures in both T2V and V2T settings, demonstrating its robustness and generalizability. PAU~\cite{pau} exhibits the largest R@1 improvement, increasing from 48.5\% to 55.4\%, indicating that our loss effectively enhances the discriminative capacity of uncertainty-adaptive models. Furthermore, X-CLIP~\cite{xclip} achieves both the highest absolute performance and the most substantial overall gain, suggesting that models incorporating multi-grained matching mechanisms are particularly well-suited for multi-event text-to-long-video retrieval.

\subsection{Ablation Study and Analysis}

To validate both the characteristics of the MELON dataset and the effectiveness of the proposed loss function, we perform a series of ablation studies and analyses. Unless otherwise specified, all experiments in this section use the popular retrieval model CLIP4Clip~\cite{clip4clip} equipped with MEA loss.

\noindent\textbf{Impact of Content Complexity on Retrieval Performance.} Analyzing retrieval performance across different video lengths and event counts provides insight into how well the dataset captures realistic variations in content complexity. As shown in \cref{fig:timestamp_event_comparison}(a), the model performs best on short and medium-length videos ($<10$ minutes), where thematic boundaries are typically cleaner and more localized. In contrast, performance drops sharply for extra-long videos, indicating that increased content diversity and narrative spread significantly elevate the difficulty of multi-event retrieval. \cref{fig:timestamp_event_comparison}(b) reveals that performance does not simply decline as the number of events increases. Instead, the model achieves its highest R@1 at four events. This suggests that when too few events are provided (\eg, $K=3$), the discriminative signal may be insufficient for distinguishing full matches from partial matches. While increasing the number of events does not guarantee improved performance, these results indicate that a moderate level of event richness provides clearer cues for long-video retrieval.

\begin{table}[t]
\centering
\begin{minipage}[t]{0.585\linewidth}
\centering
\vspace{0pt}  %
\includegraphics[width=\linewidth]{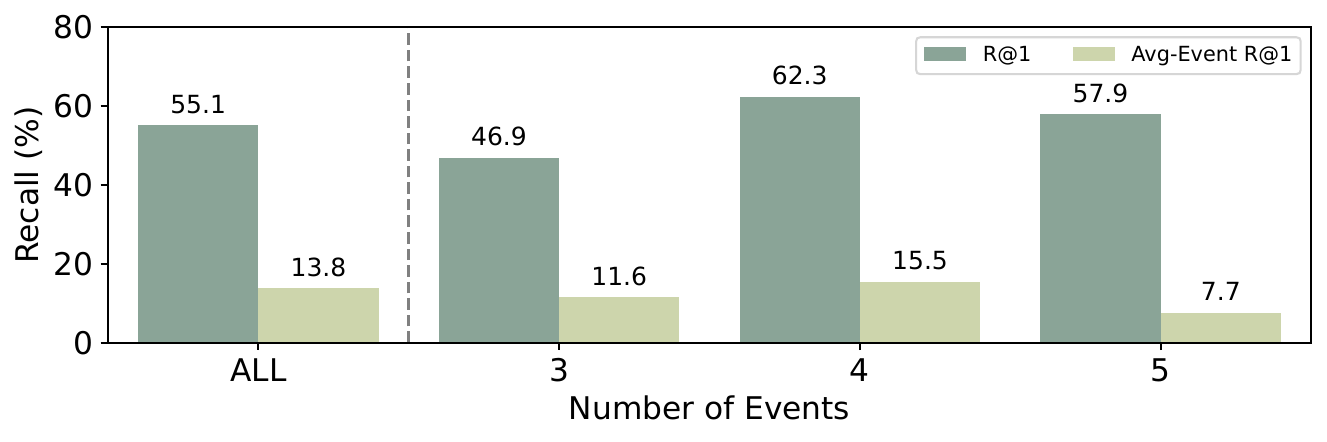}
\captionof{figure}{Difference between Text-to-Video R@1 and Avg-Event R@1 by Number of Events.}
\label{fig:event_comparison}
\end{minipage}
\hfill
\begin{minipage}[t]{0.39\linewidth}
\centering
\vspace{0pt}  %
\caption{Ablation study on MEA loss components: Evaluation of pushing term $\mathcal{L}_{push}$, upper bound $\xi^t$, and pulling term $\mathcal{L}_{pull}$.}
\label{tab:loss_element_ablation}
\resizebox{0.85\linewidth}{!}{%
\begin{tabular}{ccccccccc}
\toprule[1.5pt]
\multicolumn{3}{c}{$\boldsymbol{\mathcal{L}_{mea}}$}      & \multicolumn{6}{c}{\textbf{Text-to-Video retrieval}} \\ \cmidrule(lr){1-3} \cmidrule(l){4-9}
$\mathcal{L}_{push}$ & $\xi^t$ & $\mathcal{L}_{pull}$ & R@1 & R@5 & R@10 & Rsum & MdR & MnR \\ \midrule
 &  & \multicolumn{1}{c|}{} & 52.7 & 77.1 & 84.7 & 214.5 & 1.0 & 10.7 \\
\checkmark &  & \multicolumn{1}{c|}{} & 45.2 & 70.1 & 79.5 & 194.8 & 2.0 & 17.3 \\
\checkmark & \checkmark & \multicolumn{1}{c|}{} & 52.2 & 76.5 & 84.2 & 212.9 & 1.0 & 11.2 \\
\checkmark &  & \multicolumn{1}{c|}{\checkmark} & 52.5 & 76.6 & 85.8 & 214.9 & 1.0 & 10.7 \\
\rowcolor{lightgray}
\checkmark & \checkmark & \multicolumn{1}{c|}{\checkmark} & \textbf{55.1} & \textbf{78.1} & \textbf{86.0} & \textbf{219.2} & \textbf{1.0} & \textbf{10.7} \\ \bottomrule[1.5pt]
\end{tabular}%
}
\end{minipage}
\end{table}

\begin{table}[t]
\centering
\caption{Generalization capability of the MEA loss on standard multi-event datasets. Retrieval performance on ActivityNet-Cap and DiDeMo.}
\label{tab:generalization}
\resizebox{0.8\linewidth}{!}{%
\begin{tabular}{@{}lcccccccccccc@{}}
\toprule[1.5pt]
 & \multicolumn{6}{c}{\textbf{ActivityNet-Cap}} & \multicolumn{6}{c}{\textbf{DiDeMo}} \\ 
\cmidrule(lr){2-7} \cmidrule(l){8-13}
\multirow{3}{*}[1.1em]{\textbf{Method}} & \multicolumn{3}{c}{\textbf{Text-to-Video}} & \multicolumn{3}{c}{\textbf{Video-to-Text}} & \multicolumn{3}{c}{\textbf{Text-to-Video}} & \multicolumn{3}{c}{\textbf{Video-to-Text}} \\ 
\cmidrule(lr){2-4} \cmidrule(lr){5-7} \cmidrule(lr){8-10} \cmidrule(l){11-13}
 & R@1 & R@5 & R@10 & R@1 & R@5 & R@10 & R@1 & R@5 & R@10 & R@1 & R@5 & R@10 \\ 
\midrule
\multicolumn{1}{l|}{CLIP4Clip} & 42.6 & 73.4 & 84.8 & 42.0 & 72.8 & 85.4 & \multicolumn{1}{|c}{43.3} & 73.4 & 82.2 & 44.5 & 72.9 & 81.7 \\
\rowcolor{lightgray}
\multicolumn{1}{r|}{\ \textbf{+$\mathcal{L}_{mea}$}} & \textbf{43.8} & \textbf{75.2} & \textbf{86.1} & \textbf{45.5} & \textbf{75.8} & \textbf{87.4} & \multicolumn{1}{|c}{\textbf{45.0}} & \textbf{73.6} & \textbf{82.3} & \textbf{46.0} & \textbf{73.7} & \textbf{82.8} \\ 
\multicolumn{1}{l|}{TempMe} & 47.1 & 75.4 & 86.7 & 46.6 & 75.5 & 86.6 & \multicolumn{1}{|c}{52.7} & \textbf{81.3} & 87.9 & 49.3 & 79.6 & 87.9 \\
\rowcolor{lightgray}
\multicolumn{1}{r|}{\ \textbf{+$\mathcal{L}_{mea}$}} & \textbf{49.5} & \textbf{78.2} & \textbf{88.2} & \textbf{48.7} & \textbf{78.8} & \textbf{89.2} & \multicolumn{1}{|c}{\textbf{53.3}} & 81.0 & \textbf{88.1} & \textbf{52.5} & \textbf{81.0} & \textbf{88.6} \\ 
\bottomrule[1.5pt]
\end{tabular}%
}
\end{table}

\noindent\textbf{Evaluative Capability of MELON Test Set in Multi-Event Scenario.} To verify whether the MELON test set effectively distinguishes full multi-event matches from partial event matches, we introduce Avg-Event R@1, an extension of the standard R@1 metric. Avg-Event R@1 computes the average R@1 obtained when each single event from the ground-truth query is used independently as a retrieval query. This allows us to measure a model’s single-event retrieval ability within a multi-event evaluation environment. As shown in \cref{fig:event_comparison}, a substantial performance gap persists between the standard multi-event R@1 and Avg-Event R@1 across all event counts. This disparity is particularly pronounced for five-event queries: while the full multi-event query achieves an R@1 of 57.9\%, the Avg-Event R@1 drops sharply to 7.7\%. Such a dramatic difference demonstrates that the MELON test set inherently discourages retrieval systems from relying on partial-event matches, instead requiring a comprehensive understanding of the entire multi-event query to identify the correct video.

\noindent\textbf{Synergistic Contribution of MEA Loss Components.} We analyze how each component of the MEA loss contributes to the effective regulation of partial queries through a sequential ablation study (\cref{tab:loss_element_ablation}). Applying only $\mathcal{L}_{push}$ (Row 1) leads to severe performance degradation, indicating that pushing partial queries away from full queries without a corrective mechanism destabilizes training and disrupts the intended embedding geometry. Performance improves significantly when either the upper bound constraint $\xi^t$ (Row 2) or the $\mathcal{L}_{pull}$ term (Row 3) is introduced. The $\xi^t$ term prevents aggressive penalization by establishing a reasonable margin, while $\mathcal{L}_{\text{pull}}$ selectively restores the semantic positioning of overly pushed partial queries. The full MEA loss (Row 4), which integrates all components, achieves the best overall performance with a +2.4\% R@1 improvement over the baseline. This confirms that the components operate synergistically to maintain partial queries as true partial matches, resulting in a more coherent and semantically meaningful embedding space. Additional analyses are provided in Supplementary B.

\noindent\textbf{Generalization Capability.} To verify the generalizability of the MEA loss, we evaluate its performance on ActivityNet-Cap and DiDeMo—standard datasets featuring multi-event scenarios. As shown in \cref{tab:generalization}, integrating MEA consistently improves both Text-to-Video and Video-to-Text retrieval across different backbones. Specifically, with state-of-the-art TempMe \cite{tempme} backbone, MEA yields R@1 gains of up to +2.4\% on ActivityNet-Cap and +3.2\% on DiDeMo. These results confirm MEA as a versatile, architecture-agnostic objective that effectively regulates event-level correspondences beyond long-form scenarios.

\begin{figure}[t]
    \centering
    \includegraphics[width=0.9\linewidth]{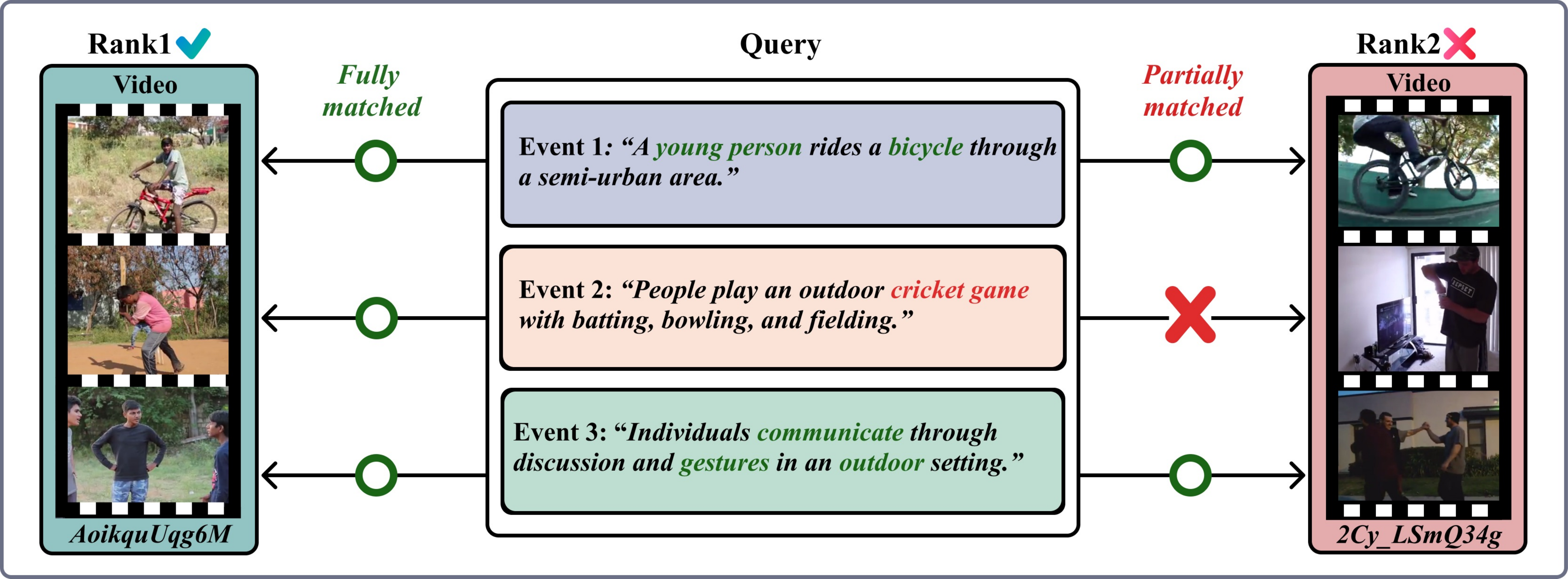}
    \caption{
    Qualitative results of text-video retrieval on the MELON test set. Our model correctly identifies the complete multiple event in the query (left) by distinguishing it from a misleading partial overlap (right). }
    
    \label{fig:qualitative_study}
\end{figure}

\subsection{Qualitative Study}

We conduct a qualitative comparison of retrieved rankings using the MELON test set. As shown in \cref{fig:qualitative_study}, the video ranked 2nd (right) is highly relevant because two of the queried events appear within it. However, only the top-ranked video (left) provides a complete match across all events in the multi-event query. This example highlights both the importance and the inherent challenge of distinguishing partially overlapping matches in long, multi-event videos. These results further reinforce the need for retrieval systems capable of accurately identifying full-event matches rather than relying on partial overlaps. Additional qualitative examples are provided in Supplementary B.

\section{Conclusion}
In this work, we introduced MELON, a large-scale dataset that successfully expands the scope of text-to-video retrieval to realistic long-form settings by incorporating complex, multi-event structures. MELON contains over 12K long-form videos paired with multiple event-level annotations and corresponding text queries, enabling realistic evaluation of retrieval in complex, multi-event scenarios. We additionally propose the MEA loss, which significantly improves retrieval accuracy by explicitly addressing the partial-event matching problem in long-form video retrieval. We anticipate that the MELON dataset and the proposed loss function, will provide a strong foundation for advancing text-video retrieval.

\bibliographystyle{splncs04}
\bibliography{main}

\setcounter{page}{1}

\setcounter{section}{0} %

\setcounter{figure}{7}
\setcounter{table}{4}

\renewcommand{\thesection}{\Alph{section}}

\noindent The supplementary material is organized as follows:

\begin{itemize}
\item We discuss the limitations of our proposed dataset and modeling in \cref{sec:sup:limitations}.
\item Detailed implementation and additional ablation studies in \cref{sec:sup:additional_experiments}.
\item The details of dataset filtering, pre-processing, quality filtering, and verification are provided in \cref{sec:sup:dataset_filtering}.
\item Further details on the dataset, including additional statistics and samples, in \cref{sec:sup:detailed_dataset}.
\item Visual samples from MELON and qualitative results of text-video retrieval are presented in \cref{sec:sup:samples_and_qualitative}. 
\item The detailed prompting strategies used for description, summarization, and judgment in \cref{sec:sup:detailed_prompt}.
\end{itemize}

\section{Limitations} %
\label{sec:sup:limitations}
While MELON and the proposed MEA loss establish a strong baseline for multi-event long-video retrieval, several limitations remain. First, due to the inherent distribution of the source data, the dataset contains a relatively small number of extra-long videos ($>$30 min). This imbalance may lead to underfitting during training specifically for these extreme duration cases 
(shown in Fig. 5(a)).
Second, a structural limitation arises when the number of query events $K$ becomes excessively large. As discussed in \cref{sec:sup:ablation}, this necessitates a modification to the encoding strategy. Future work could address this by adopting sophisticated encoder architectures capable of handling longer textual contexts.

\section{Additional Experiments} %
\label{sec:sup:additional_experiments}
\subsection{Implementation Details} We detail the common implementation settings that apply to the seven representative models \cite{clip4clip, xclip, ts2net, ucofia, uatvr, pau, tempme} and the specific configurations for our proposed MEA loss.

\subsubsection{Training Settings.} For all six models, the maximum token length for text input is set to 248, following the configuration of LongCLIP~\cite{longclip}. We utilize the Adam~\cite{adam} optimizer, and configure the initial learning rate for the LongCLIP module to 1e-7 and 1e-4 for other modules. This is accompanied by a cosine scheduling strategy and a warm-up proportion of 0.1. All models are trained for 10 epochs with a batch size of 16. Any other training settings not explicitly mentioned here adhere to the default configurations of each respective model.

\subsubsection{MEA Loss Parameters.} For constituting the partial query set within the MEA loss, the number of events $m$ excluded from the original query is set to 2, and the size of the partial query set $L$ is fixed at 3. The $\lambda$ values for the MEA loss are individually tuned for each model to achieve optimal performance, with the corresponding ablation study detailed in \cref{sec:sup:ablation} and \cref{fig:lambda_ablation}. The optimized $\lambda$ values for each model are: CLIP4Clip (0.2), X-CLIP (0.7), TS2-Net (0.1), UCOFIA (0.7), UATVR (0.2), PAU (0.3), and TempMe (0.2).

\subsection{Ablation Study}
\label{sec:sup:ablation}
This section presents additional ablation studies to further evaluate our proposed method. Unless otherwise noted, all ablation studies are conducted on the CLIP4Clip~\cite{clip4clip} model in this section.

\subsubsection{Query Encoding Type.} 
The choice of query encoding strategy is critical for a model's ability to comprehend long contextual information. We primarily employ a common concatenation-based method, which involves joining all event sentences into a single sequence before encoding to retain comprehensive contextual information. However, a practical limitation arises with this approach: the maximum token length an encoder can process. For instance, our current LongCLIP-based encoder is constrained by a maximum token length of 248 tokens. As the number of query events $K$ increases, directly concatenating all event sentences can exceed this limit, posing a structural challenge and implicitly necessitating a modification to the encoding strategy.

To investigate alternative strategies, we also examine an approach where each event sentence is individually encoded, followed by a mean pooling operation. For a fair comparison, this analysis was conducted without integrating the MEA loss. \cref{tab:query_type} demonstrates that the mean pooling-based encoding strategy generally achieves lower performance compared to our concatenation approach. Specifically, in terms of R@1, the pooling method achieves 49.7\%, which is 3.0\% lower than the concatenation method. This outcome indicates that while pooling can circumvent length limitations, the independent processing of event sentences struggles to capture the holistic context, leading to reduced performance. This suggests that the lack of global tokenization and self-attention across the entire query explicitly limits the model's ability to fully grasp the long narrative context.
\begin{table}[ht]
\caption{Text-to-Video Retrieval performance differences based on the query encoding type.}
\label{tab:query_type}
\centering
\resizebox{0.6\columnwidth}{!}{%
\begin{tabular}{ccccccc}
\toprule[1.5pt]
\multirow{2}{*}{\textbf{Query type}} & \multicolumn{6}{c}{\textbf{Text-to-Video retrieval}} \\ \cmidrule(l){2-7} 
 & R@1    & R@5    & R@10    & Rsum    & MdR    & MnR   \\ \midrule
\multicolumn{1}{l|}{Event concatenation} & \textbf{52.7} & \textbf{77.1} & \textbf{84.7} & \textbf{214.5} & \textbf{1.0} & \textbf{10.7} \\
\multicolumn{1}{l|}{Event pooling} & 49.7 & 74.6 & 83.1 & 207.4 & 2.0 & 12.6 \\
\bottomrule[1.5pt]
\end{tabular}%
}
\end{table}

\subsubsection{Ablation on the MEA Loss Weight ($\lambda$).}
\begin{figure*}[!ht]
    \centering
    \includegraphics[width=1.0\linewidth]{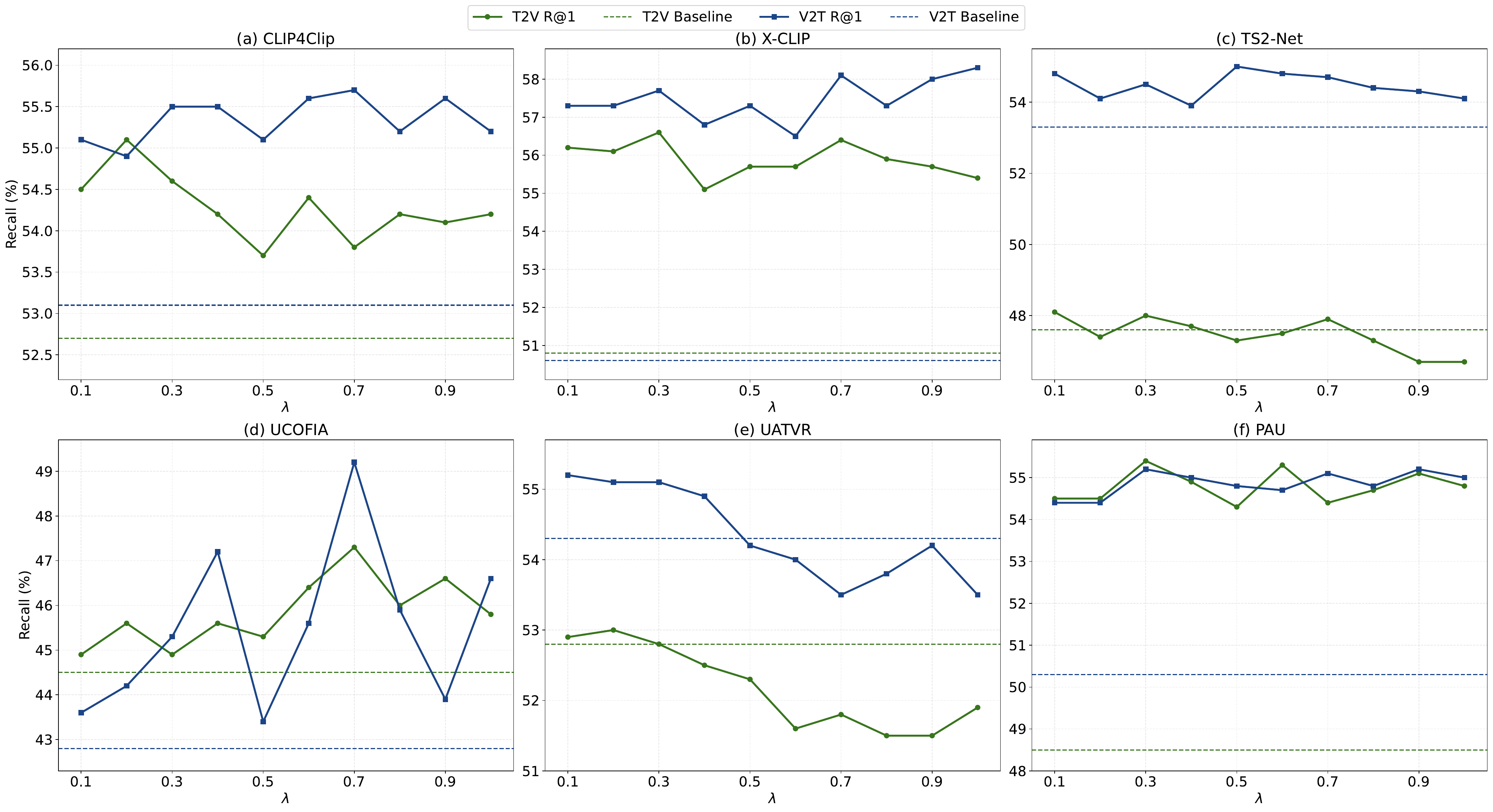}
    \caption{Ablation study on $\lambda$ (Weight of MEA Loss). The effect of varying the weight $\lambda \in [0.1, 1.0]$ on the performance of six representative text-video retrieval models. Each subplot illustrates the T2V R@1 and V2T R@1 measured on the MELON dataset. The T2V and V2T baselines (i.e., performance when $\lambda=0$) are marked by the dashed lines in their respective colors.}
    \label{fig:lambda_ablation}
\end{figure*}

To assess the robustness of the proposed MEA loss, we conduct validation experiments by varying its weight ratio $\lambda$ across multiple retrieval models. As shown in \cref{fig:lambda_ablation}, incorporating the MEA loss generally yields clear and stable retrieval performance gains across the entire range of $\lambda$ values for most models. Notably, the performance improvement is often more pronounced in V2T retrieval. With the exception of UATVR~\cite{uatvr} and TS2-Net~\cite{ts2net}, all other evaluated models consistently demonstrate these stable gains, indicating that the MEA loss is effective without requiring sensitive tuning of its combination ratio.

\subsubsection{Impact of Individual MEA Loss Components on Training Dynamics.}
\begin{figure}[!ht]
    \centering
    \includegraphics[width=1.0\linewidth]{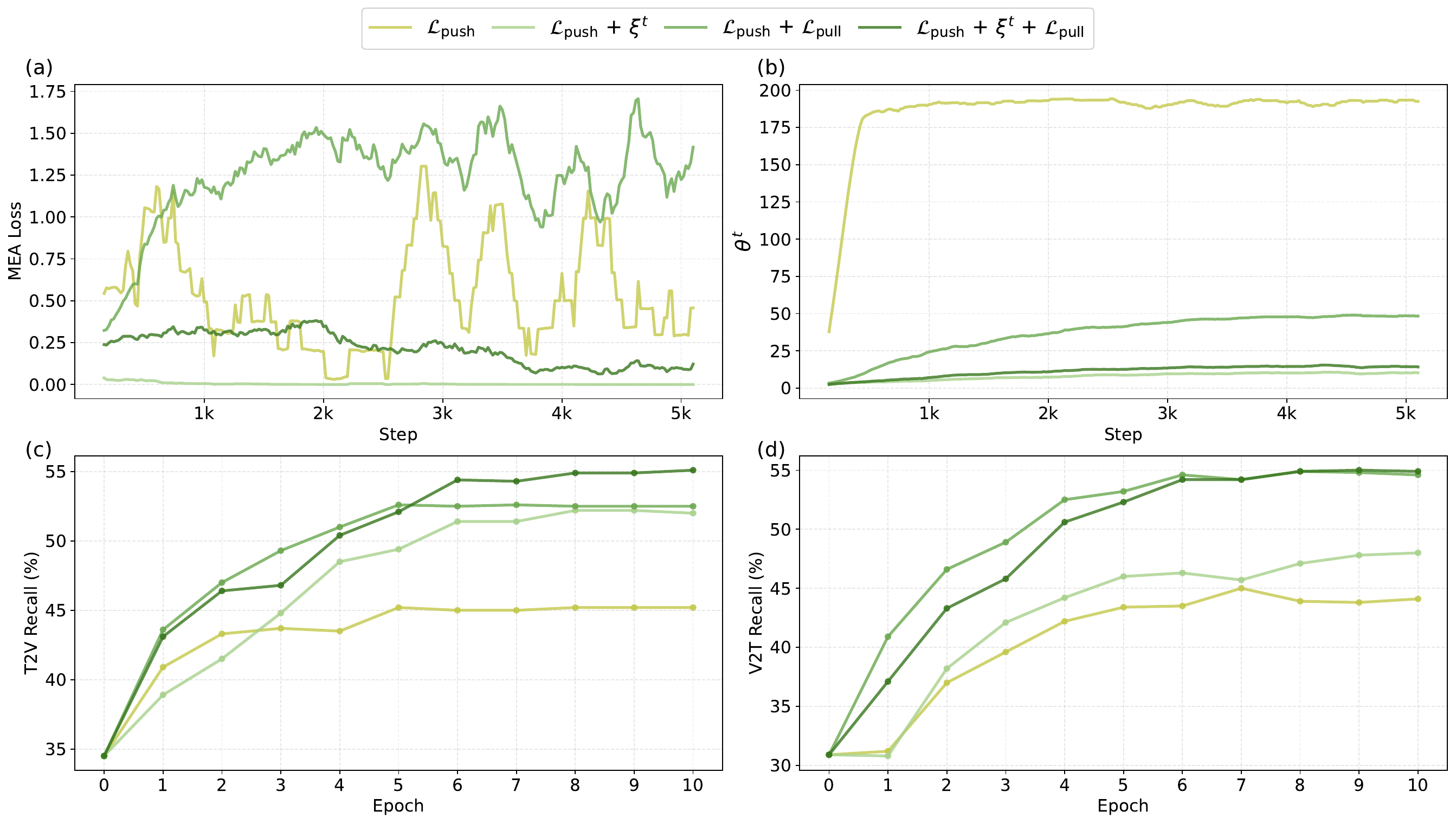}
    
    \caption{Ablation study on the training dynamics and retrieval performance impacted by the individual MEA loss components ($\mathcal{L}_{\mathrm{push}}$, $\xi^t$, $\mathcal{L}_{\mathrm{pull}}$). The four panels show: (a) the training loss, (b) the adaptive margin threshold $\theta^t$ over steps, and (c, d) the corresponding T2V and V2T R@1 performance over epochs.}
    \label{fig:mce_training_ablation}
\end{figure}

In addition to the final performance impact of loss components (Tab. 4), \cref{fig:mce_training_ablation} visually represents how these components influence the curriculum learning approach and subsequent performance changes over training. Specifically, the figure details: (a) the training loss dynamics, (b) the evolution of the adaptive margin threshold $\theta^t$ across training steps, and (c, d) the T2V and V2T R@1 performance, respectively, over epochs.

When only the $\mathcal{L}_{push}$ component is utilized, we observe highly erratic training loss dynamics (\cref{fig:mce_training_ablation}(a)) and a consistently high threshold $\theta^t$ (\cref{fig:mce_training_ablation}(b)) throughout the training steps. These results support the argument that merely pushing without regulation leads to unstable learning and distorts the intended embedding geometry. Consequently, \cref{fig:mce_training_ablation}(c) and \cref{fig:mce_training_ablation}(d) reveal that this configuration yields the lowest retrieval performance for both T2V and V2T.

The inclusion of the upper-bound constraint $\xi^t$ alongside $\mathcal{L}_{push}$ shows only a marginal impact on the training dynamics when examining \cref{fig:mce_training_ablation}(a) and \cref{fig:mce_training_ablation}(b). Despite the added constraint, V2T performance remains low (\cref{fig:mce_training_ablation}(d)). This indicates that while bounding the push might prevent some extremes, further mechanisms are necessary for proper alignment of partial queries within the embedding space.

Conversely, the addition of the $\mathcal{L}_{pull}$ component to $\mathcal{L}_{push}$ results in a substantial performance improvement, particularly prominent in the V2T task (\cref{fig:mce_training_ablation}(d)). This highlights the critical role of the pull term in correcting overly pushed embeddings or addressing cases where the model incorrectly evaluates certain events, which is essential for achieving proper alignment. Although training dynamics in \cref{fig:mce_training_ablation}(a) and \cref{fig:mce_training_ablation}(b) are more stable than with $\mathcal{L}_{push}$ alone, some instability persists.

Finally, incorporating all components of the MEA loss not only leads to remarkably stable learning (\cref{fig:mce_training_ablation}(a) and \cref{fig:mce_training_ablation}(b)) but also achieves the best performance for both T2V and V2T retrieval (\cref{fig:mce_training_ablation}(c) and \cref{fig:mce_training_ablation}(d)). This conclusively demonstrates that each component plays a crucial role in effectively aligning partial queries, contributing synergistically to the overall objective. As a result, the MEA loss significantly enables the model to distinguish subtle event-level differences in multi-event scenarios.

\subsubsection{Robustness to Query Event Ordering.}
In realistic scenarios, users often formulate multi-event queries without strict temporal constraints. Ideally, while a query with the correct event ordering should yield the highest similarity, queries describing the same events in a different order should still achieve comparable similarity. To evaluate whether our model exhibits this property, we test its performance under reversed and shuffled query event orderings. As shown in \cref{tab:mea_ordering}, our method experiences only a marginal performance drop compared to the default ordering, demonstrating that the learned representations are not overly sensitive to the order in which events are described.
\begin{table}[!ht]
\centering
\caption{Robustness of MEA loss to variations in query event ordering.}
\label{tab:mea_ordering}
\resizebox{0.65\columnwidth}{!}{%
\begin{tabular}{@{}lcccccc@{}}
\toprule[1.5pt]
\multirow{2}{*}{\textbf{Query Event Order}} & \multicolumn{3}{c}{\textbf{Text-to-Video}} & \multicolumn{3}{c}{\textbf{Video-to-Text}} \\ \cmidrule(lr){2-4} \cmidrule(l){5-7}
 & R@1 & R@5 & R@10 & R@1 & R@5 & R@10 \\ \midrule
Reversed & 53.1 & 76.5 & 84.5 & 53.3 & 77.8 & 86.0 \\
Shuffled & 53.6 & 76.6 & 84.9 & 53.6 & 78.2 & 85.4 \\
\textbf{Default} & \textbf{55.1} & \textbf{78.1} & \textbf{86.0} & \textbf{54.9} & \textbf{79.3} & \textbf{86.9} \\
\bottomrule[1.5pt]
\end{tabular}
}%
\end{table}

\subsubsection{Partial Queries vs.\ Text Augmentation: Does Event-Specific Information Matter?}
To validate the necessity of our partial query strategy, we compare it against standard text augmentations (e.g., word replacement and event shuffling) as an alternative source of partial supervision in the MEA loss. One might expect that such augmentations could serve as a proxy for partial queries, since a shuffled or word-replaced sentence still retains only a subset of the original semantics. However, as reported in \cref{tab:mea_augmentation}, replacing partial queries with these augmentations degrades performance below the baseline, suggesting that generic textual perturbations lack the information necessary for the model to distinguish between full and partial event coverage and may rather hinder than help the training signal. These results support the use of explicitly constructed partial queries for effective multi-event learning.

\begin{table}[!ht]
\centering
\caption{Comparison of MEA's partial queries against text augmentation strategies.}
\label{tab:mea_augmentation}
\resizebox{0.7\columnwidth}{!}{%
\begin{tabular}{@{}lcccccc@{}}
\toprule[1.5pt]
\multirow{2}{*}{\textbf{Method}} & \multicolumn{3}{c}{\textbf{Text-to-Video}} & \multicolumn{3}{c}{\textbf{Video-to-Text}} \\ \cmidrule(lr){2-4} \cmidrule(l){5-7}
 & R@1 & R@5 & R@10 & R@1 & R@5 & R@10 \\ \midrule
CLIP4Clip & 52.7 & 77.1 & 84.7 & 53.1 & 76.9 & 85.0 \\
\ + Text augmentation & 48.5 & 72.7 & 81.3 & 36.1 & 61.9 & 71.9 \\
\textbf{\ + Partial queries (Ours)} & \textbf{55.1} & \textbf{78.1} & \textbf{86.0} & \textbf{54.9} & \textbf{79.3} & \textbf{86.9} \\
\bottomrule[1.5pt]
\end{tabular}
}%
\end{table}

\section{Dataset Filtering} %
\label{sec:sup:dataset_filtering}
In this section, we provide a comprehensive breakdown of our multi-stage filtering pipeline. This process is critical to transforming raw, noisy web-sourced videos into a high-fidelity benchmark suitable for multi-event retrieval.

\subsection{Preprocessing}
To construct a dataset that genuinely reflects \textit{long-form} and \textit{multi-event} characteristics, we initially integrate data from six prominent video datasets \cite{cgbench, youcook2, longvale, scenewalk, vidchapter7m, vrbench}. While these sources provide a valuable foundation, they are not immediately usable for our specific task due to variable annotation quality.
As noted in the main paper, web-sourced datasets (e.g., subsets from YouTube) frequently suffer from issues such as broken URLs, metadata misalignment, or overly sparse captions. To mitigate this, we apply a strict preprocessing protocol:
\begin{enumerate}
    \item \textbf{Duration Constraint:} We strictly remove videos shorter than 2 minutes to ensure sufficient temporal capacity for multiple events.
    \item \textbf{Annotation Completeness:} We discard samples lacking essential timestamp information or containing broken text descriptions.
\end{enumerate}
As shown in \cref{tab:dataset_preprocess}, this initial cleansing significantly reduces noise. For instance, from an initial pool of $2,007$ candidates in `youcook2`, we retain $1,690$ high-quality samples, ensuring that all subsequent processing steps are applied to valid data.

\begin{table}[!ht]
\caption{Dataset Processing Statistics}
\label{tab:dataset_preprocess}
\centering
\resizebox{0.8\columnwidth}{!}{%
\begin{tabular}{lrrr}
\toprule[1.5pt]
\textbf{Dataset} & \textbf{Raw Videos} & \textbf{Preprocessed Videos} & \textbf{Ratio (\%)} \\
\midrule
VR-Bench & 960 & 924 & 96.25\% \\
VideoChapter7M & 817,076 & 6,547 & 0.80\% \\
CG-Bench & 1,219 & 1,218 & 99.92\% \\
LongVALE & 8,411 & 6,402 & 76.11\% \\
Youcook2 & 2,007 & 1,690 & 84.21\% \\
Scenewalk & 87,514 & 6,000 & 6.86\% \\
\midrule
\textbf{Total} & \textbf{917,187} & \textbf{22,781} & \textbf{2.48\%} \\
\bottomrule[1.5pt]
\end{tabular}}

\end{table}

\subsection{Quality Filtering}
After preprocessing and initial caption generation, we employ a dual-filtering mechanism to ensure both visual relevance and semantic distinctiveness.

\begin{table}[!ht]
\caption{Dataset filtering statistics. The threshold for each dataset was determined empirically.}
\label{tab:v2t_filterling}
\centering
\small
\begin{tabular}{lcrrr}
\toprule[1.5pt]
\textbf{Dataset} & \textbf{Threshold} & \textbf{Total} & \textbf{Kept (\%)} \\
\midrule
cg-bench        & 0.12 & 11{,}678   & 96.4 \\
longvale        & 0.12 & 105{,}730  & 99.9 \\
scenewalk       & 0.12 & 45{,}424   & 96.9 \\
videochapter7m  & 0.18 & 34{,}130   & 14.6 \\
scenewalk       & 0.12 & 45{,}424   & 85.2 \\
vrbench         & 0.12 & 11{,}668   & 95.5 \\
youcook2        & 0.12 & 13{,}829   & 99.6 \\
\bottomrule[1.5pt]
\end{tabular}
\end{table}

\subsubsection{V2T (Video-to-Text) filtering} 
To filter out hallucinations or descriptions that do not align with the visual content, we measure the semantic similarity between the video segment and its generated text. We utilize \cite{v2t} to compute similarity scores. 

\cref{tab:v2t_filterling} summarizes the statistics of the V2T filtering process. We empirically determined the filtering threshold for each dataset by manually inspecting 100 samples. As shown in the table, datasets known for noisy web-scraped annotations, such as \texttt{videochapter7m}, showed a significant reduction (retaining only 14.6\%), indicating the necessity of this filtering step. Conversely, manually curated datasets like \texttt{longvale} and \texttt{youcook2} maintained very high retention rates (99\%), confirming their initial high alignment quality.

\subsubsection{T2T (Text-to-Text) Filtering.} 
To ensure the \textit{multi-event} nature of the dataset, we remove videos where consecutive events are semantically redundant (e.g., a continuous shot of a lecture with repetitive captions). We calculate the cosine similarity between adjacent event descriptions using SBERT.
\begin{itemize}
    \item \textbf{Criterion:} If the similarity between adjacent captions exceeds 0.70, the video is flagged as having low event diversity and is subsequently removed. This step ensures that each video in MELON contains distinct, non-overlapping narrative events.
\end{itemize}

\subsection{Human Verification and LMM Judging}
For the test set, we implement a rigorous verification pipeline combining a Large Multimodal Model (LMM)-based judge with human review.

\subsubsection{LMM-Based Screening.}
We utilize Gemini 2.5-flash to audit all test set summaries. The model is prompted to identify inconsistencies across the triplet: \{\textit{Video}, \textit{Timestamp}, \textit{Description}\}. Out of the 3,600 test videos, the model flags 617 instances (17.14\%) as potentially containing at least one inaccurate summary.

\subsubsection{Human Review Protocol.}
The flagged videos are subsequently examined by three expert annotators. They utilize a custom UI displaying the video segment alongside the generated summary and classify the error type (e.g., \textit{Hallucination}, \textit{Vague Description}).
\begin{itemize}
    \item \textbf{Consensus:} The annotators achieve an Inter-Annotator Agreement (IAA) of 86.2\%. Any remaining disagreements are resolved through discussion to reach a final consensus regarding whether to modify or remove the problematic segments.
\end{itemize}

\subsubsection{Agreement Distribution.}
To provide deeper insight into the Inter-Annotator Agreement (IAA), \cref{tab:agreement_dist} details the distribution of consensus levels among the three annotators. Out of the 617 flagged videos, 532 samples (86.2\%) received a unanimous decision (either all `Keep' or all `Modify' / `Remove'), demonstrating the robustness of our annotation guidelines.
\begin{table}[ht]
    \centering
    \caption{Distribution of Annotator Agreement for the 617 Flagged Samples. `Unanimous' indicates all three annotators made the same decision, showing high reliability.}
    \resizebox{0.85\columnwidth}{!}{
    \begin{tabular}{lccc}
        \toprule[1.5pt]
        \textbf{Agreement Level} & \textbf{Vote Pattern (Keep:Modify)} & \textbf{Count} & \textbf{Ratio (\%)} \\
        \midrule
        \multirow{2}{*}{\textbf{Unanimous (3/3)}} & 3 : 0 & 98 & 15.9 \\
                                                  & 0 : 3 & 434 & 70.3 \\
        \midrule
        \multirow{2}{*}{\textbf{Majority (2/3)}}  & 2 : 1 & 26 & 4.2 \\
                                                  & 1 : 2 & 59 & 9.6 \\
        \midrule
        \textbf{Total} & - & \textbf{617} & \textbf{100.0} \\
        \bottomrule[1.5pt]
    \end{tabular}
    }

    \label{tab:agreement_dist}
\end{table}

\subsubsection{Reliability Check (Addressing false negatives).}
To validate the reliability of the LMM judge and ensure that the ``unflagged'' portion (approx.\ 83\%) is indeed clean, we perform a spot-check on a random sample of 100 unflagged videos. Human review confirms that 97\% of these samples are error-free, demonstrating that the LMM judge has a high recall for detecting errors and justifying our focus on the flagged subset.

\section{Detailed Dataset} %
\label{sec:sup:detailed_dataset}
\begin{figure}[ht]    
    \label{fig:appendix_theme_dur}
    \centering
    \includegraphics[width=\linewidth]{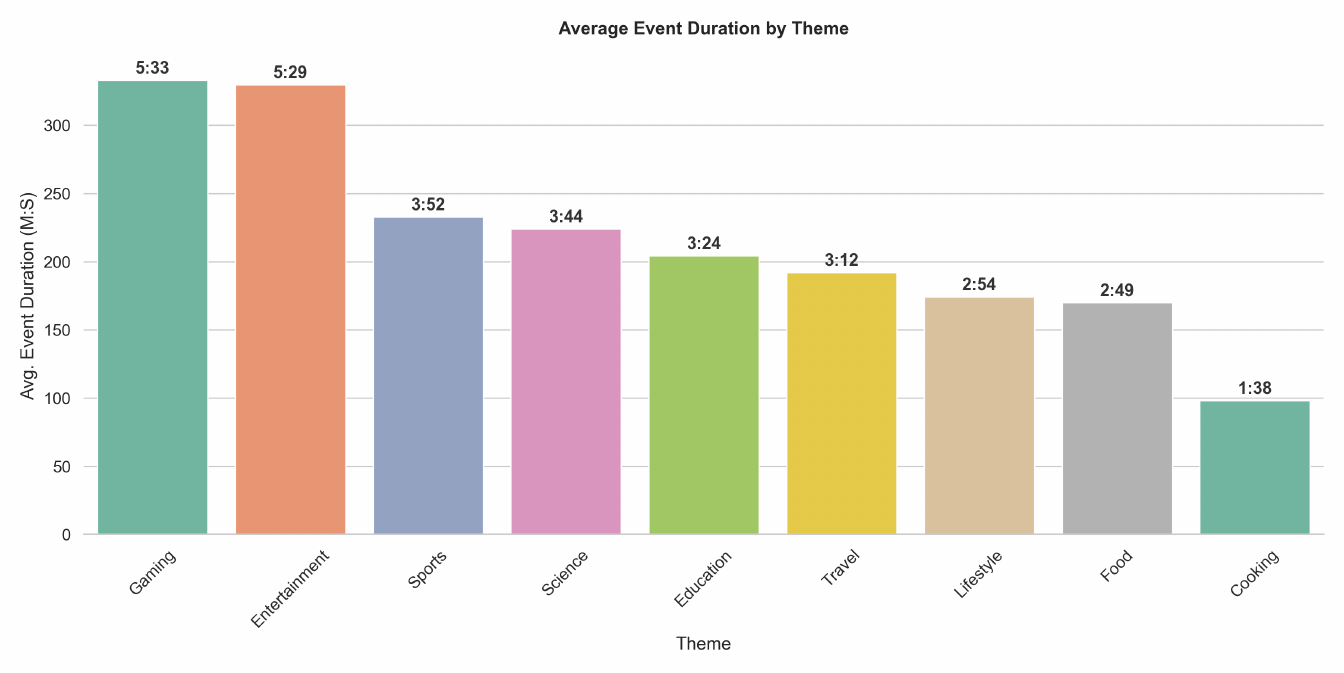}
    \caption{Distribution of average event duration by theme.}
\end{figure}
\begin{figure}[ht]    
    \centering
    \includegraphics[width=\linewidth]{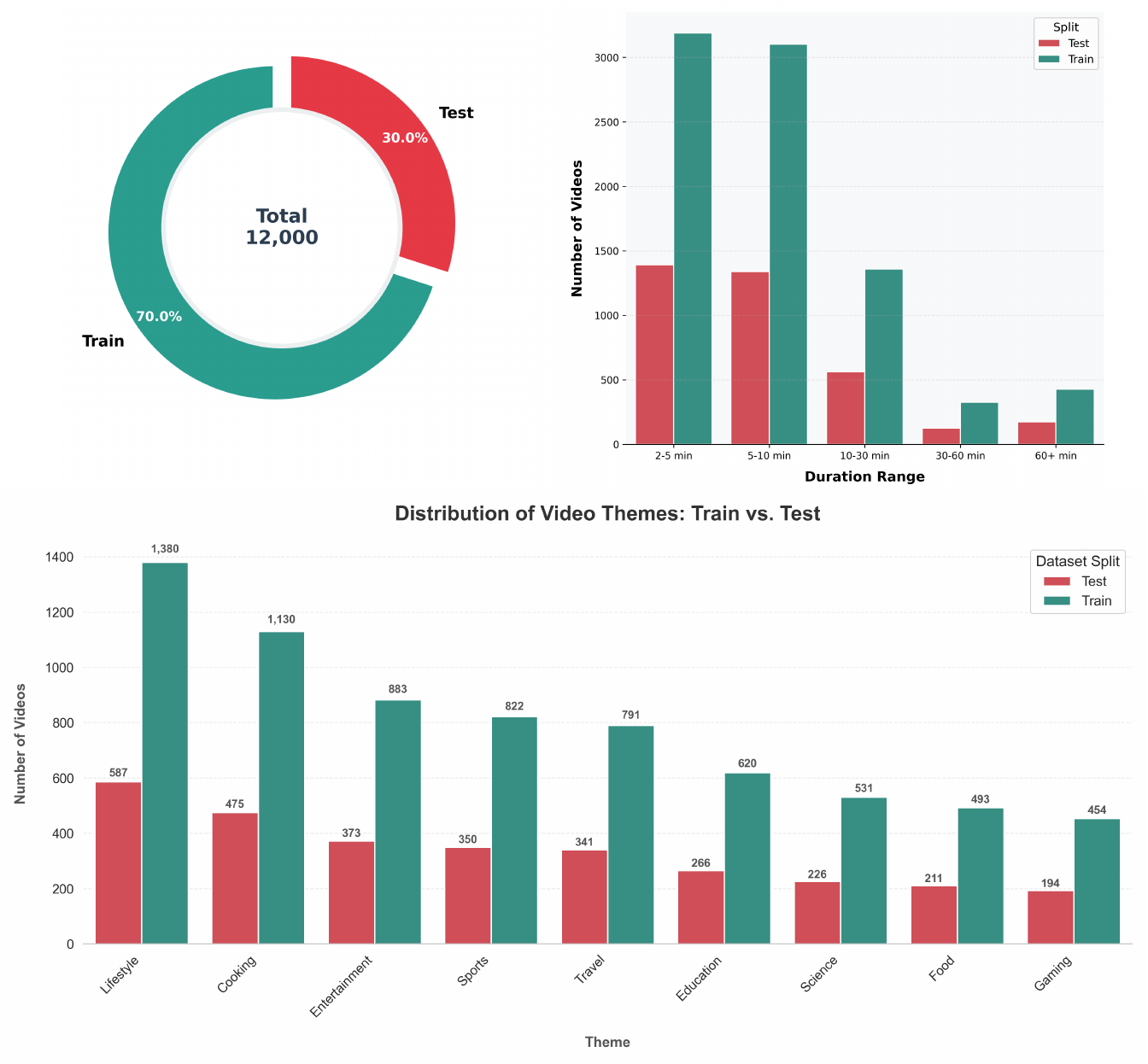}
    \caption{\textbf{Distribution of the Train and Test sets.} Illustrates the data split ratio, duration ranges, and video counts per theme. The charts demonstrate a balanced distribution between training and testing subsets across all categories and durations.}
\label{fig:dataset_train_test}
\end{figure}

In this section, we analyze the statistics of the MELON dataset to underscore its suitability for long-form video understanding.

We begin by examining the temporal extent of events across various themes, as shown in \cref{fig:appendix_theme_dur}. Unlike standard datasets that typically focus on short, atomic actions, MELON features events that span several minutes—for instance, Gaming and Entertainment events average over 5 minutes. This extended duration allows the dataset to better capture the complexity found in real-world videos.

To ensure a robust and reliable evaluation, we adopt a balanced splitting strategy illustrated in \cref{fig:dataset_train_test}. While maintaining a 70:30 ratio (8,400 training and 3,600 test samples), we carefully preserve the distribution of video durations and themes. This approach guarantees that the test set remains a representative and unbiased sample of the whole.

Beyond statistical distributions, we also visualize the semantic richness of the dataset in \cref{fig:theme_wordclouds}. The distinct vocabulary observed across major themes indicates that MELON covers a wide range of topics, which helps minimize domain overfitting. Further qualitative examples in \cref{fig:sample_cooking,fig:sample_gaming,fig:sample_travel,fig:sample_education} demonstrate the density and contextual relevance of our multi-event annotations.

\clearpage %
\section{Dataset Samples and Qualitative Results on MELON} \label{sec:sup:samples_and_qualitative}

In this section, we provide visual examples of the MELON dataset's richness and the qualitative retrieval result of our proposed method. 

\subsection{MELON Dataset Samples}

\begin{figure*}[ht]    
    \centering
    \includegraphics[width=\textwidth]{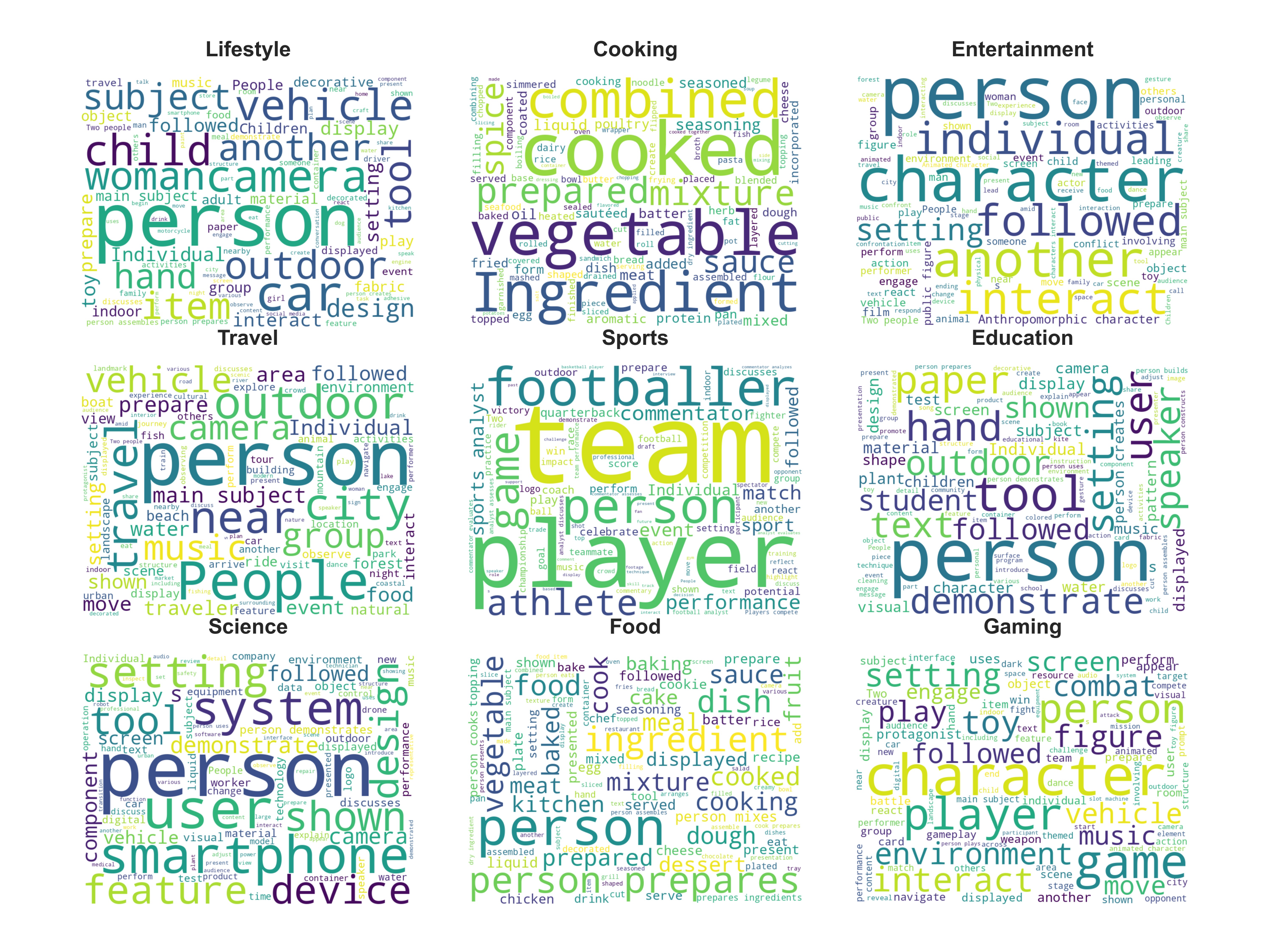}
    \caption{\textbf{Word cloud visualization by major themes.} This illustrates the distribution of frequent terms within each theme, highlighting the lexical diversity and thematic distinctiveness of the MELON dataset.}
\label{fig:theme_wordclouds}
\end{figure*}

\begin{figure*}[ht]    
    \centering
    \includegraphics[width=\textwidth]{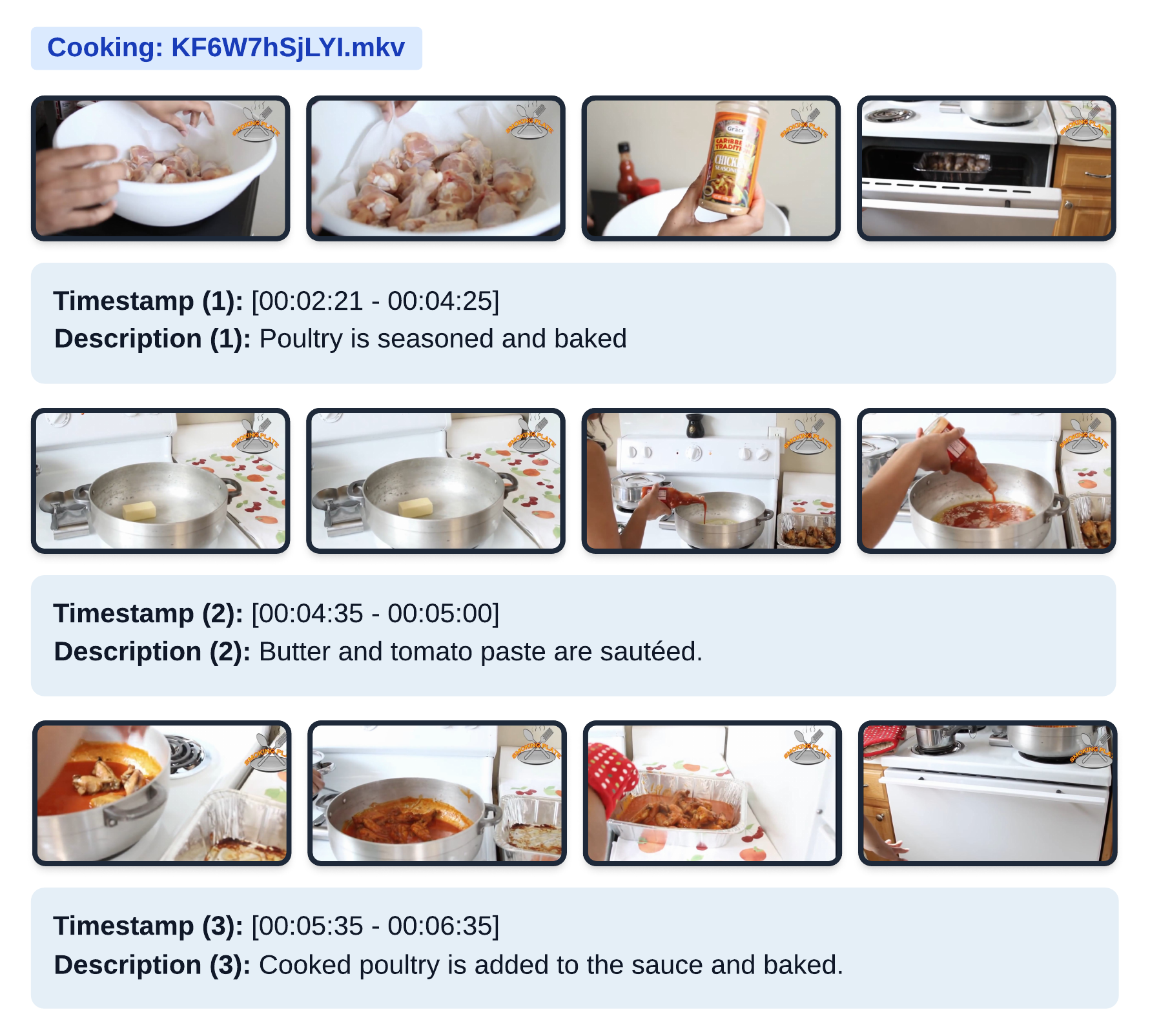}
    \caption{The \textbf{cooking theme} example from the MELON dataset, presenting the chronological events with timestamp and description for preparing a baked poultry dish with tomato sauce (involving seasoning, sautéing sauce base, and baking).}
\label{fig:sample_cooking}
\end{figure*}
\begin{figure*}[ht]    
    \centering
    \includegraphics[width=\textwidth]{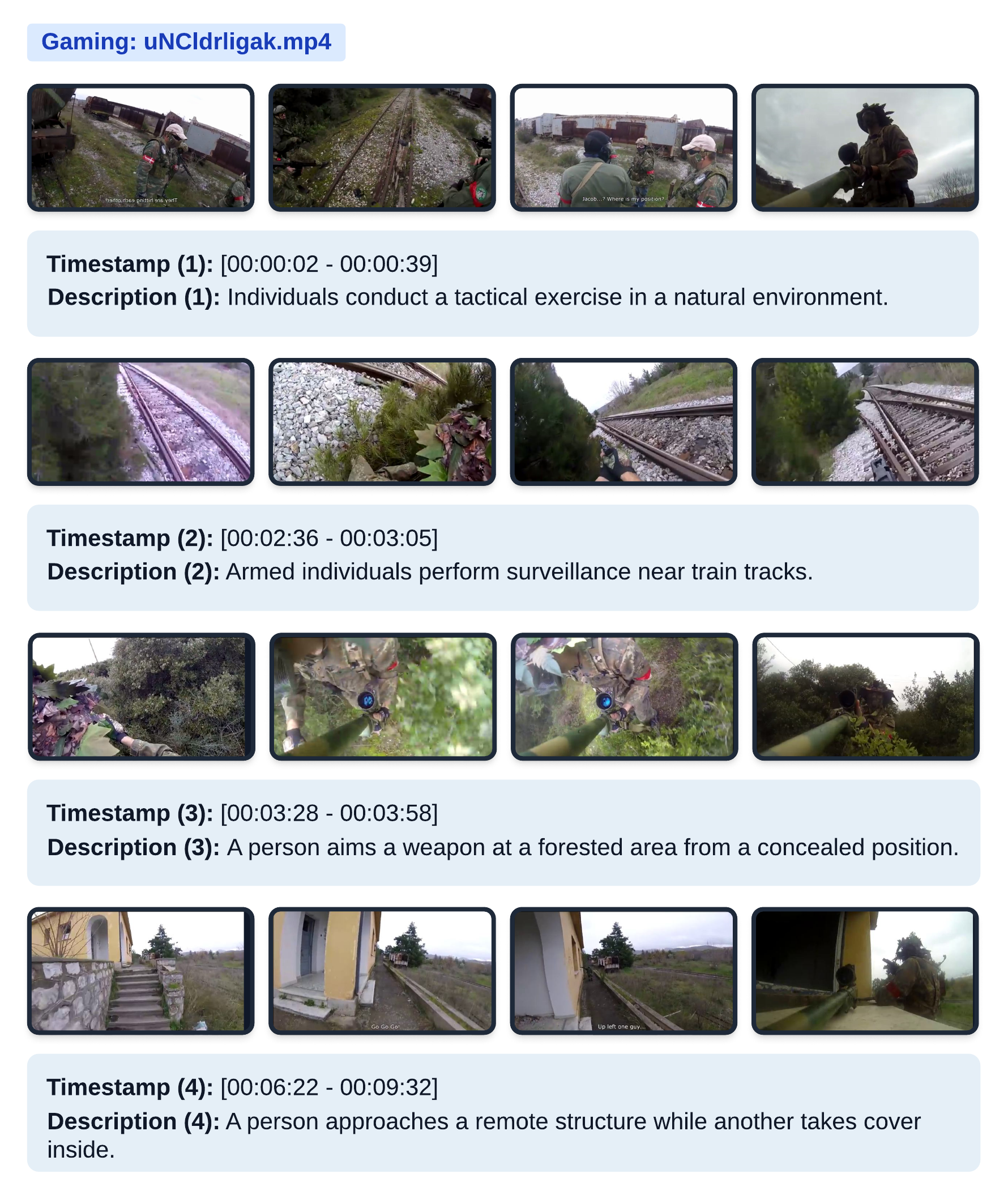}
    \caption{The \textbf{gaming theme} example from the MELON dataset, presenting the chronological events with timestamp and description for a tactical exercise in a natural environment (involving surveillance near train tracks, aiming weapons, and approaching structures).}
\label{fig:sample_gaming}
\end{figure*}
\begin{figure*}[ht]    
    \centering
    \includegraphics[width=0.98\textwidth]{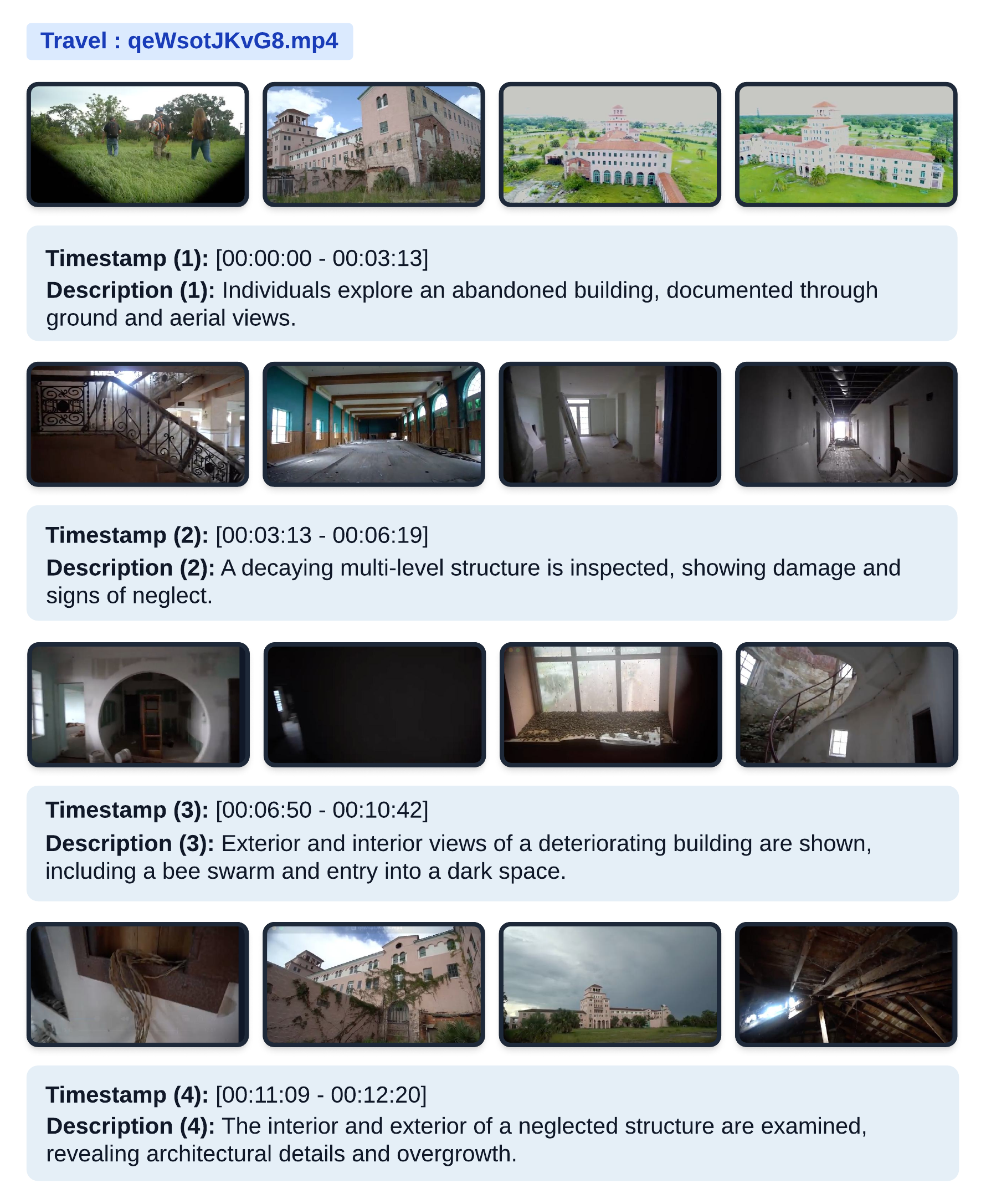}
    \caption{The \textbf{travel theme} example from the MELON dataset, presenting the chronological events with timestamp and description for exploring an abandoned and deteriorating structure (involving ground and aerial views, examining interior damage, and observing exterior architectural details).}
\label{fig:sample_travel}
\end{figure*}
\begin{figure*}[ht]    
    \centering
    \includegraphics[width=\textwidth]{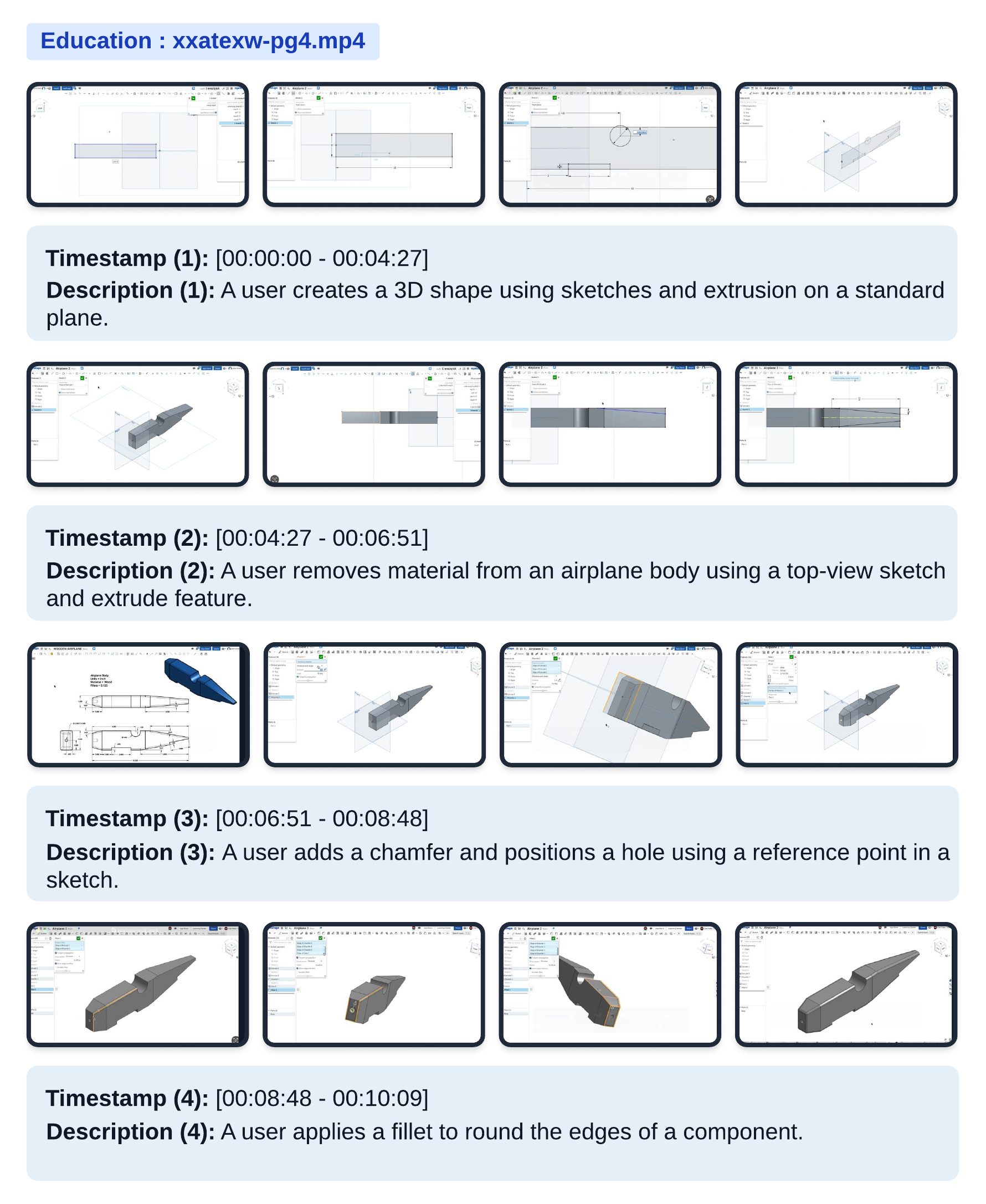}
    \caption{The \textbf{education theme} example from the MELON dataset, presenting the chronological events with timestamp and description for a 3D modeling tutorial (involving creating shapes using extrusion, removing material, adding chamfers and fillets to a component).}
\label{fig:sample_education}
\end{figure*}

We visualize the lexical diversity of the MELON dataset through word clouds for major themes in \cref{fig:theme_wordclouds}. Furthermore, we present representative examples from the Cooking, Gaming, Travel, and Education themes (\cref{fig:sample_cooking,fig:sample_gaming,fig:sample_travel,fig:sample_education})to illustrate the high-fidelity alignment between multi-event video segments and their corresponding descriptions.

\clearpage
\subsection{Qualitative Results}
\begin{figure*}[ht]    
    \centering
    \begin{subfigure}{\textwidth}
        \centering
        \includegraphics[width=\textwidth]{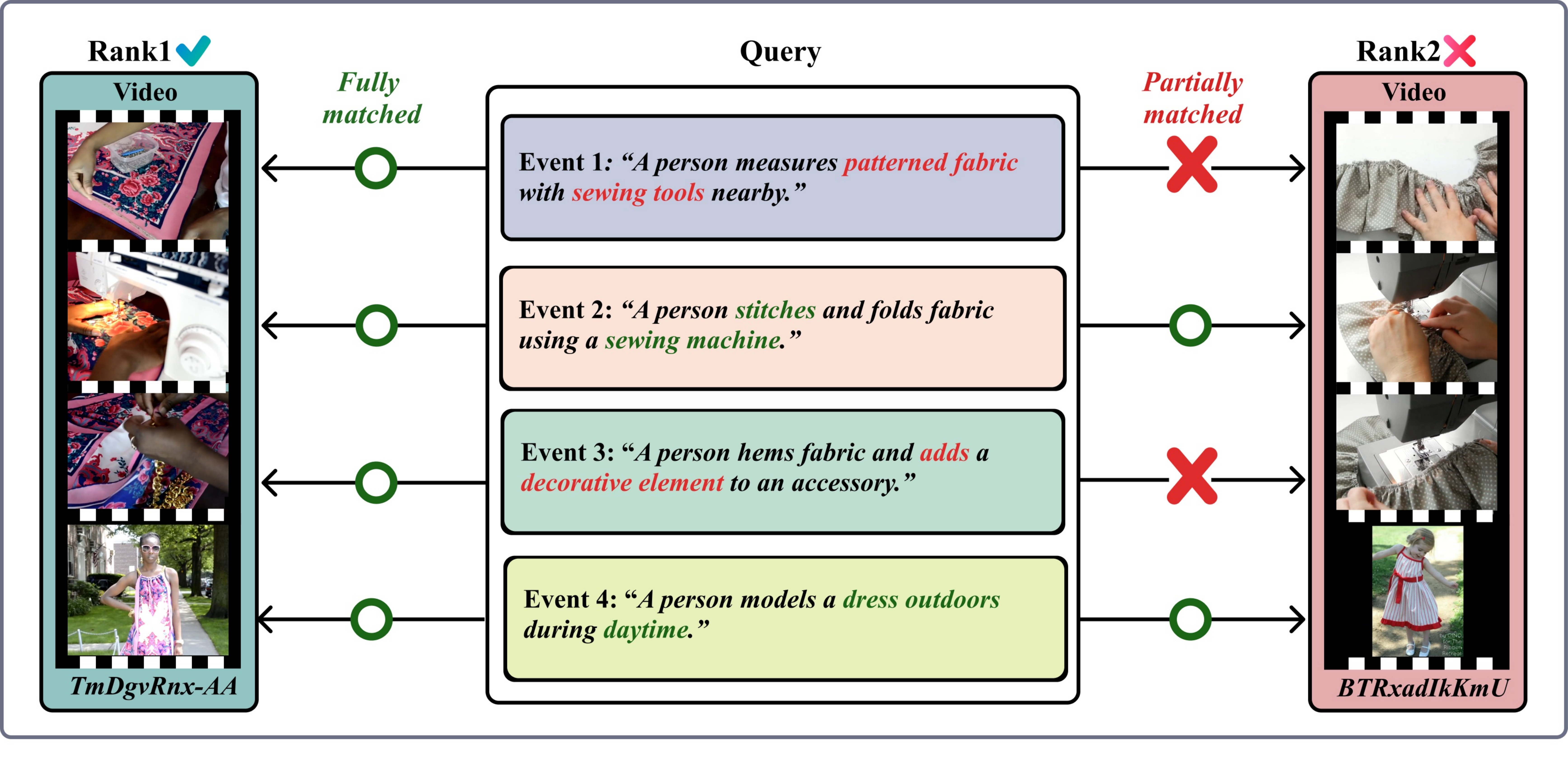}
        \caption{Sewing and modeling}
        \label{fig:retrieval_sample1}
    \end{subfigure}
    
    \begin{subfigure}{\textwidth}
        \centering
        \includegraphics[width=\textwidth]{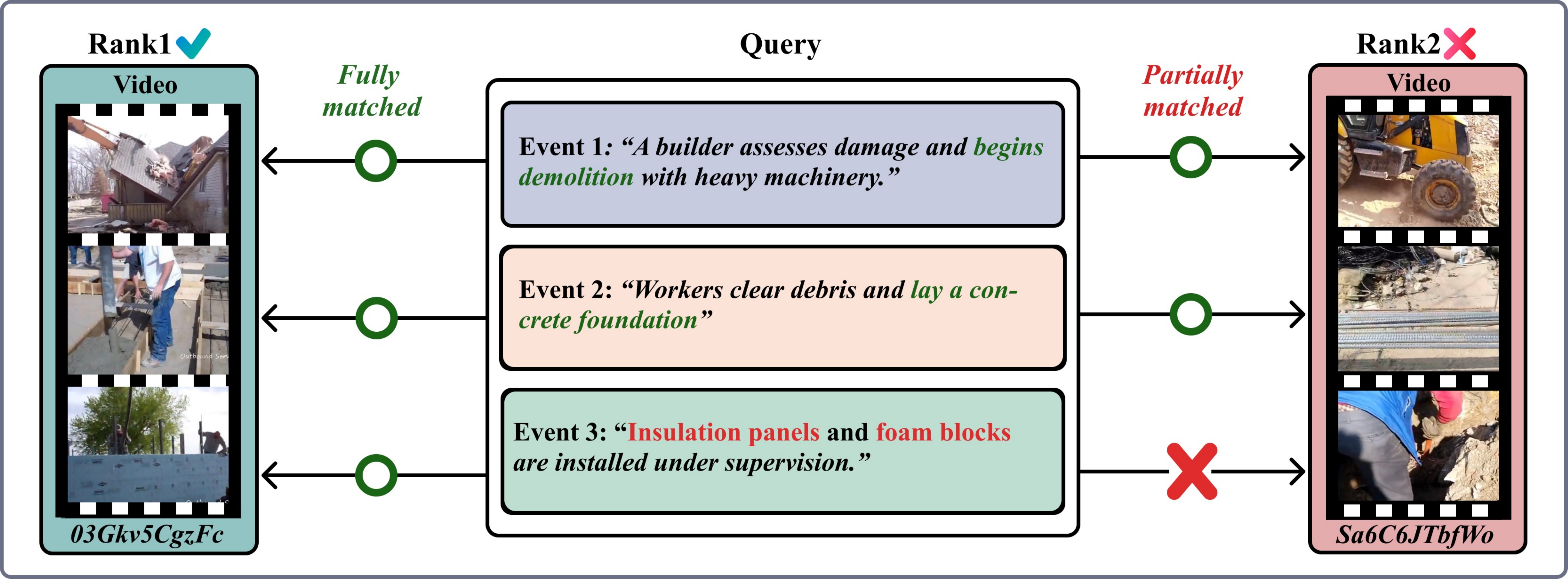}
        \caption{Construction activities}
        \label{fig:retrieval_sample2}
    \end{subfigure}
    
    \caption{\textbf{Qualitative results of text-video retrieval.} Our model correctly identifies the complete multiple events in the query (left) by distinguishing them from a misleading partial overlap (right).}
    \label{fig:retrieval_sample_12}
\end{figure*}
To further demonstrate the effectiveness of our proposed method, we provide qualitative retrieval results. As illustrated in \cref{fig:retrieval_sample_12}, our model successfully discriminates the fully matched video from visually similar but partially matched videos by accurately capturing fine-grained semantic details across multiple events. Specifically, as shown in \cref{fig:retrieval_sample1}, even when a Rank 2 video contains visually similar actions such as 'stitching,' our model correctly identifies the absence of other key events such as 'measuring' or 'hemming' to maintain the correct ranking. A similar trend is observed in the construction scenario (\cref{fig:retrieval_sample2}), and these results underscore the robustness of the MEA loss in distinguishing subtle semantic differences within complex video contexts.

\clearpage
\section{Detailed Prompt} %
\label{sec:sup:detailed_prompt}

For reproducibility and to support future research, we present the exact prompts used in each stage of our data construction and evaluation pipeline. The design of each prompt category is visually summarized in \cref{fig:qa_prompt}, \cref{fig:multievent_prompt}, and \cref{fig:lmm_judge_prompt}.

\subsection{Prompt for Generating Dense Descriptions}
\begin{figure*}[ht]    
    \centering
    \includegraphics[width=\textwidth]{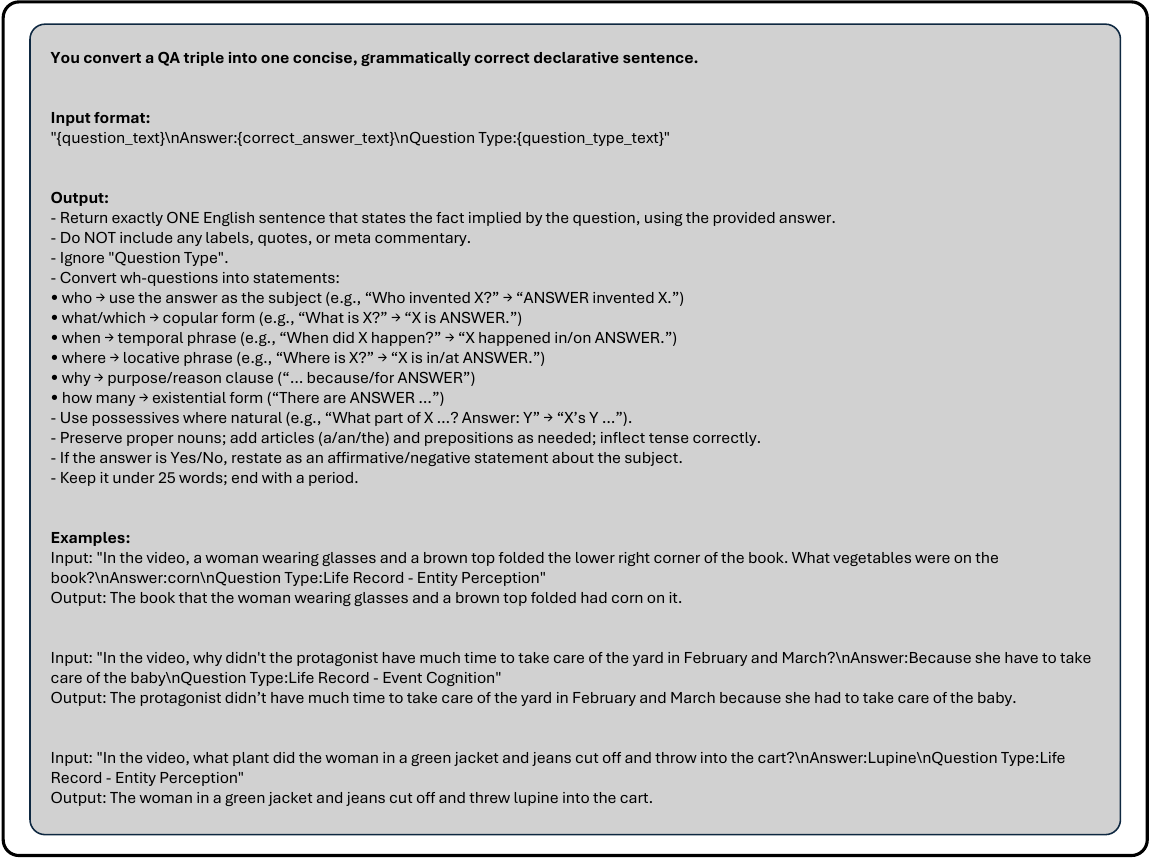}
    \caption{\textbf{Prompt for QA-to-Caption conversion.} This illustrates the rules and examples for synthesizing answers into natural sentences based on question types (wh-questions, yes/no, etc.)}
\label{fig:qa_prompt}
\end{figure*}

We convert raw QA pairs or sparse metadata into dense, retrieval-oriented captions using the prompt summarized in \cref{fig:qa_prompt}.  
The figure highlights the instruction schema that rewrites short answers into fluent, information-rich descriptions based on question type (e.g., wh-questions, yes/no).  

\subsection{Prompt for Generating Multi-Event Summaries}
\begin{figure*}[ht]    
    \centering
    \includegraphics[width=\textwidth]{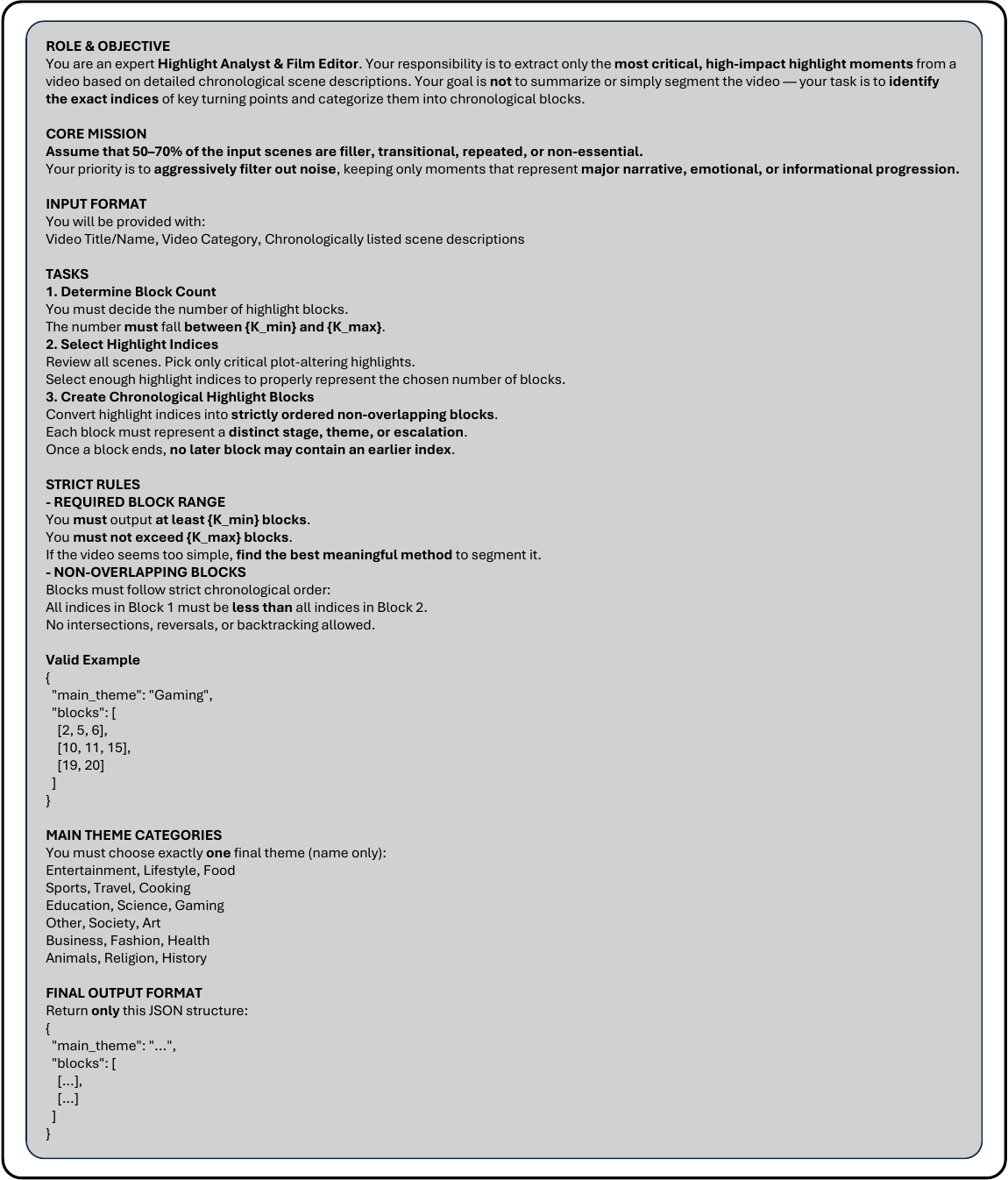}
    \caption{\textbf{Prompt for multi-event summary generation.} This illustrates the instructions guiding the LLM to filter noise from input scene descriptions and select only narrative-rich highlight segments, outputting them in a structured JSON format.}
\label{fig:multievent_prompt}
\end{figure*}

To produce high-level narratives from fine-grained scene segments, we utilize the multi-event summarization prompt shown in \cref{fig:multievent_prompt}.  
This prompt instructs the model to suppress noisy or irrelevant segments and retain only semantically meaningful highlights.  
It also enforces a structured JSON format containing \textit{Theme}, \textit{Description}, and \textit{Timestamp}.  

\subsection{Prompt for LMM Judge}
\begin{figure*}[ht]    
    \centering
    \includegraphics[width=\textwidth]{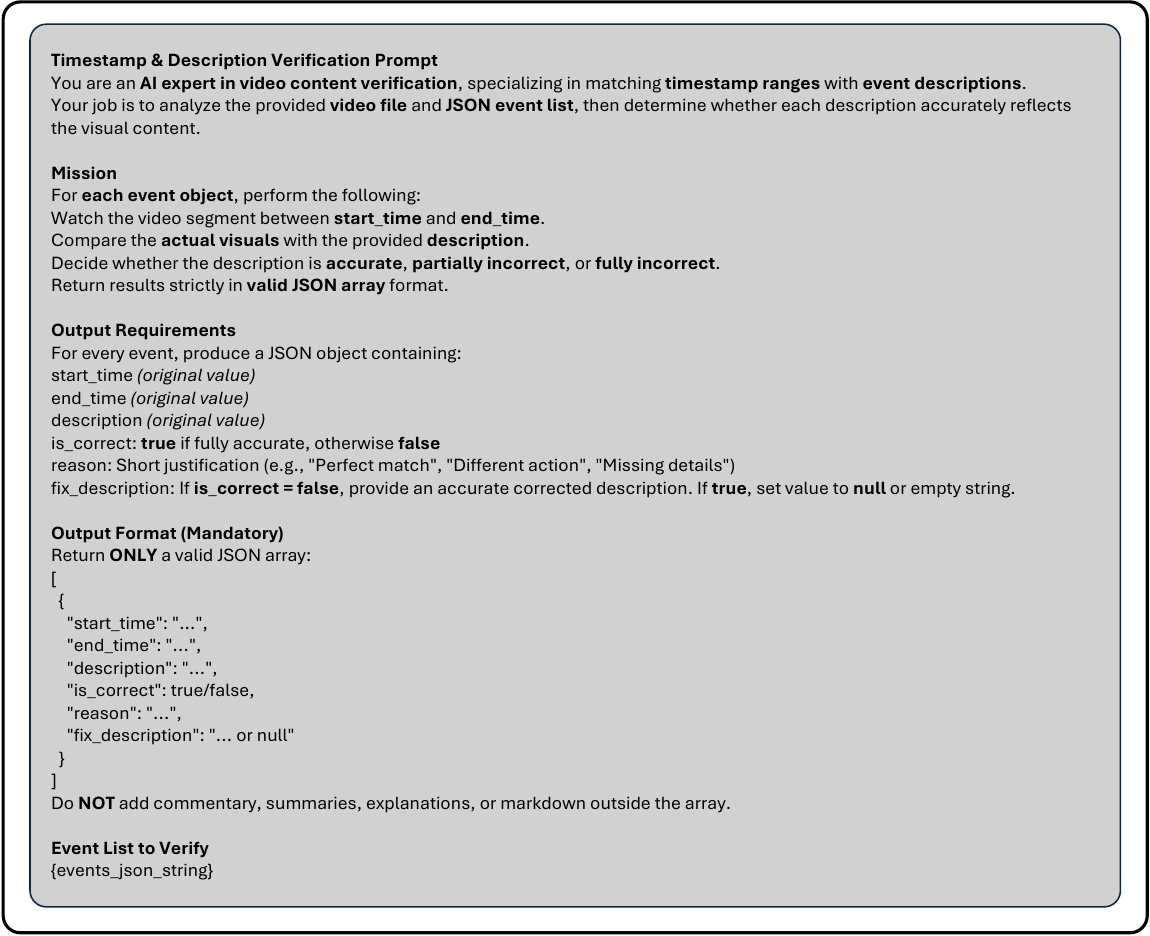}
    \caption{\textbf{LMM instructions for timestamp and description verification.} A prompt designed to automatically judge the alignment between visual content and event descriptions, directing the LMM to analyze the video and return specific reasons and correction suggestions for inaccurate segments.}
\label{fig:lmm_judge_prompt}
\end{figure*}

For automatic quality verification, we employ the LMM-judge prompt illustrated in \cref{fig:lmm_judge_prompt}.  
The prompt guides the model to evaluate the alignment between visual evidence and the associated textual descriptions and timestamps.  
It further requires the model to identify specific inconsistencies and provide targeted correction suggestions prior to human inspection.  

\subsection{Human Filtering Stage}
\begin{figure*}[ht]    
    \centering
    \includegraphics[width=\textwidth]{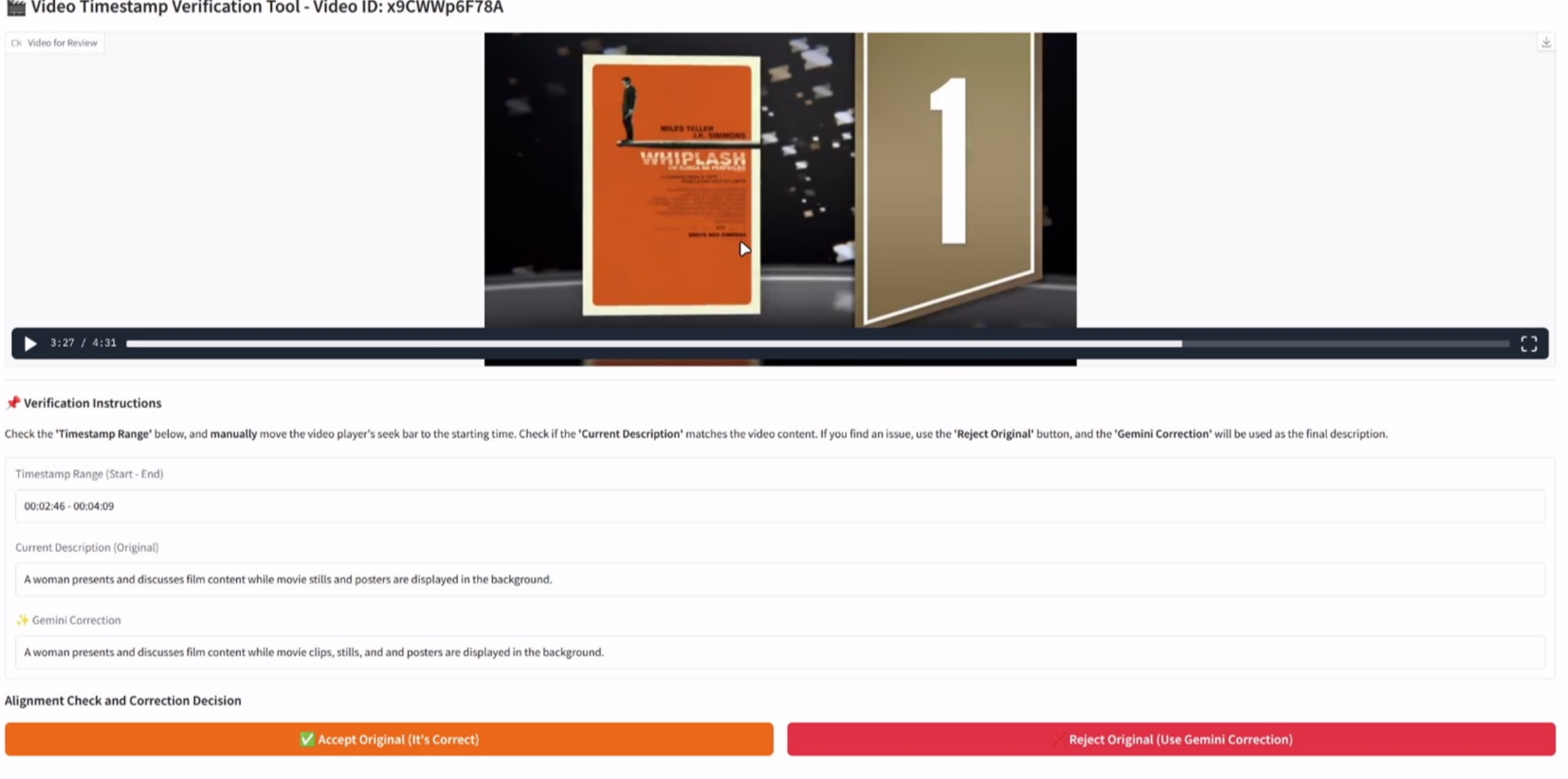}
    \caption{\textbf{Interface used during the human review stage.} Annotators evaluate the coherence of the \{Video, Description, Timestamp\} triplet and classify issues such as hallucination or vague description mismatch.}
\label{fig:human_filter}
\end{figure*}

Finally, human annotators review each \{Video, Description, Timestamp\} triplet using the interface depicted in \cref{fig:human_filter}.  
During this stage, annotators assess coherence, identify issues such as hallucination or vague mismatches, and finalize the quality labels for the dataset.  

\clearpage
\clearpage

\end{document}